\documentclass[useAMS,usenatbib]{mn2e}
\usepackage{graphicx}
\usepackage{epstopdf}
\usepackage[figuresright]{rotating}
\usepackage{subfigure,color}
\usepackage{longtable}
\usepackage{xspace}

\newcommand{\ion}[2]{{\textrm{#1}}\,{\textrm{\sc #2}}}

\title[Sulphur abundance determinations in star-forming regions]
{Sulphur abundance determinations in star-forming regions-I: Ionization Correction Factor}
\author[Dors et al.]
{O.~L. Dors Jr.$^{1}$\thanks{E-mail:\,olidors@univap.br}, E. P\'erez-Montero$^{2}$, G.~F. H\"agele $^{3,4}$, M.~V. Cardaci$^{3,4}$,  
\newauthor{A.~C. Krabbe$^{1}$}\\
$^1$ Universidade do Vale do Para\'iba, Av. Shishima Hifumi, 2911, Cep12244-000, S\~ao Jos\'e dos Campos, SP, Brazil\\
$^{2}$ Instituto de Astrof\'{\i}sica de Andaluc\'{\i}a (CSIC), PO Box 3004, E-18080 Granada, Spain\\
$^3$ Instituto de Astrof\'isica de La Plata (CONICET-UNLP), Argentina. \\
$^4$ Facultad de Ciencias Astron\'omicas y Geof\'{\i}sicas, Universidad Nacional de La Plata, Paseo del Bosque s/n, 1900 La Plata, Argentina.
}
\begin{document}

\date{Accepted- 2011 April 28. Received -2011 February 18.}

\pagerange{\pageref{firstpage}--\pageref{lastpage}} \pubyear{2011}

\maketitle

\label{firstpage}

\begin{abstract}
 
In the present work  we used a grid of  photoionization models  combined with 
stellar population synthesis models  to derive reliable Ionization Correction Factors (ICFs) for the sulphur in 
 star-forming regions.  These models cover a large range of ne\-bu\-lar parameters and yielding ionic
abundances in consonance with those derived  through optical and infrared observational data of star-forming regions. 
From our theoretical ICFs, we suggested an $\alpha$ value of $3.27\pm0.01$ in the  classical Stasi\'nska formulae. 
We compared  the total sulphur abundance in the gas phase  of a large sample of   objects 
by using our Theoretical ICF and other approaches.
In average, the differences between the determinations via the use of the different ICFs considered are similar to the  uncertainties in the S/H estimations. 
Nevertheless, we noted that for some objects it could reach up to about  0.3 dex for the low metallicity regime.
 Despite of the large scatter of the points, we found a trend of S/O ratio to decrease with the metallicity, independently of the ICF used to 
compute the sulphur total abundance.

 \end{abstract}


\section{Introduction}

The knowledge of the  abundance  of heavy elements (e.g. O, S, N, Ne)  in the gas phase 
of star-forming regions play a key role in  studies of stellar nucleosynthesis,  initial mass function of stars and  chemical
evolution of galaxies.

To derive  the total abundance of a given element (X) in ionized nebulae, after to estimate the electron temperature and
electron density of the gas phase, it is necessary  to calculate the abundance of all  its  ionization stages (see \citealt{osterbrock89}).
However, for the majority of the elements present in star-forming regions,  only    emission-lines of some  ionization stages  can be measured.
In these cases, the use of Ionization Correction Factors (ICFs) is necessary to derive the contribution of unobserved ions, as initially defined  by 
\citet{peimbert69}  

\begin{equation}
\label{eq00}
 \rm ICF(X^{+i})=\frac{X/H}{X^{+i}/H^{+}},
\end{equation}
where $\rm X^{+i}$ is the ion whose ionic abundance  can be calculated from its observed emission-lines.

In particular, for  sulphur,  in the most of the cases the total abundance is calculated by 
a direct determination of the abundance of the ions $\rm S^{+}$ and $\rm S^{2+}$,
through the lines  [\ion{S}{ii}]$\lambda$$\lambda$6716,31
and   [\ion{S}{iii}]$\lambda$$\lambda$9069, 9532 respectively, and by 
using an ICF to correct the unobserved $\rm S^{3+}$, which produces
forbidden lines at 10.51$\mu$m. In the pioneer work,   \citet{stasinska78a} 
proposed an ICF for the sulphur  based on both $\rm S^{+}$ and $\rm S^{2+}$ ions and
given by 
\begin{equation}
\label{eq0}
 \rm ICF(S^{+}+S^{2+})= \left[1-\left(1-\frac{O^{+}}{O}\right)^{\alpha}\right]^{-1/\alpha}.
\end{equation}
Along decades,  the  value of $\alpha$ have been largely discussed in the literature. For example,
\citet{stasinska78a}, using the photoionization models of \citet{stasinska78b}, which assume the 
 Non Local Thermodynamic Equilibrium (NLTE)
stellar atmosphere models of \citet{mihalas72},  
 suggested  $\alpha=3$.  \citet{french81}, who used a sample of \ion{H}{ii} regions and planetary nebulae,  
 derived  $\alpha=2$.  \citet{garnett89}   combined spectroscopic data of \ion{H}{ii} regions
 containing the  [\ion{S}{iii}]$\lambda$$\lambda$9069, 9532  emission-lines
 (not considered by most of previous works) with photoionization models assuming different stellar atmosphere models
 in order to estimate an ICF for the sulphur.
  From this analysis,  \citet{garnett89} suggested that  an intermediary $\alpha$ value   between 2 and 3 is correct. 
   \citet{vermeij02a}, using the  optical and infrared spectroscopic data  of \citet{vermeij02}, were able to derive directly an
   ICF for the sulphur and  concluded that $\alpha=3$ is correct for $\rm O^{+}/O \: > 0.2$, being  their results less clear for higher 
 ionization stages (see also \citealt{dennefeld83, izotov94, thuan95, kwitter01, kennicutt03, enrique06}).
  Direct estimations for the sulphur ICF, such as the one performed by  \citet{vermeij02a}, require infrared spectroscopic
 data of  \ion{H}{ii} regions as well as direct measures of electron temperatures, difficult for objects with low ionization
 degrees \citep{bresolin05}. Thus, sulphur ICFs have been mainly calculated  by using  photoionization mo\-dels,   
 in which not comparison with observational  data  are performed.

Other important subject is  the relative abundance between   sulphur and oxygen, which  has a direct impact on studies of stellar nucleosynthesis.  
 These elements arise from the nucleosynthesis in massive stars \citep{arnett78, woosley95}, 
however, there are two fundamental issues ill-defined: (a) The  knowledge of the mass range of stars  that dominates the production of these elements.  
(b)  If the initial mass function  (IMF) of stars is universal. 
For decades, studies based on optical spectroscopic data of  star-forming regions
have been used to solve these problems but,   not conclusive results were obtained.
 For example, \citet{garnett89},  who derived sulphur abundances for a sample of 13 extragalactic 
\ion{H}{ii} regions, 
found a constant S/O abundance  over a range of O/H (generally used as metallicity tracer),  which suggests 
that either these elements are produced   by massive stars within  a similar mass range or
by stars of different masses but  with an universal IMF \citep{henry99}.
 This result is supported by the majority of other works done in this direction (e.g. \citealt{berg13, guseva11, enrique06, kennicutt03}). 
However, evidences of S/O ranges with O/H
were found, for example, by \citet{vilchez88} in the galaxy M\,33 and by \citet{diaz91} in M\,51. 
Moreover,  due to  large dispersion   in  S/O for a  fixed  value of O/H \citep[see e.g.][]{hagele12,hagele08,hagele06},
the idea that S/O  does not range with  the metallicity is somewhat uncertain \citep{enrique06, carolina06}.
 
 In this paper, we employ a grid of photoionization mo\-dels of \ion{H}{ii} regions  and a large sample of optical and infrared spectroscopic data of star-forming regions
   with the  following goals:

1. To derive  ICFs for the sulphur based on a consistent  comparison between ionic abundances predicted by photoionization
model and   calculated from observational data.

2.  To compare the discrepancy 
in S/H abundances computed by using different ICFs. 

3. To investigate  the  S/O-O/H relation in star-forming regions considering different ICFs for the sulphur. 
 
 This paper is the first (Paper I) of a series of three works, where  in the out-coming papers 
 we will   present    a comparison of $\rm S^{2+}/H^{+}$ abundances obtained from optical and infrared lines and 
a comparison between S/O and O/H abundances with prediction of chemical evolution models.
Similar analysis was performed for the neon by \citet{dors13}.
The present paper is organized as follows. In Section~\ref{obs} the observational data used
along the paper are presented.  In Section~\ref{mod}, we describe the 
 photoionization models used to derive  ICFs for the sulphur, while 
methodology adopted to derive the ionic abundances is given in Section~\ref{ionic}.
In Section~\ref{result} the results containing the ICFs obtained by using photoionization models and 
from observational emission-lines are presented. 
 Discussion and  conclusions regarding the outcome
are given in Sections~\ref{disc} and \ref{conc}, respectively.

\section{Observational data}
\label{obs}

We compiled from the literature   emission-line intensities of \ion{H}{ii} regions and star-forming galaxies
obtained in the optical and infrared spectral ranges. These measurements were used to obtain sulphur and oxygen ionic 
abundances  in order to  verify if our photoionization models are  representative of real \ion{H}{ii} regions,
to check   if the  theoretical ICFs are compatible with the ones derived  
 directly from observations and investigating the S/O-O/H relation.
The selection criterion for the Visible-sample was the  detection of the intensity
lines   [\ion{O}{ii}]$\lambda$$\lambda$3726+ 29 (hereafter re\-fe\-reed as [\ion{O}{ii}]$\lambda$3727),  
[\ion{O}{iii}]$\lambda$$\lambda$4363, 5007, [\ion{S}{ii}]$\lambda$$\lambda$6717, 31 
 and  [\ion{S}{iii}]$\lambda$9069. In the cases where the [\ion{S}{ii}]$\lambda$6717 and  $\lambda$6731 lines were not
resolved, the sum of the intensity of these lines were considered. 
 For some objects (indicated in Table~\ref{tab1}) the theoretical relation 
$I$[\ion{S}{iii}]$\lambda$9069=$I$[\ion{S}{iii}]$\lambda$9532/2.5 
was used to estimate the emission line intensity of  $\lambda$9069,
since only the sum of these was available.

\begin{figure}
\centering
\includegraphics[angle=-90,width=9cm]{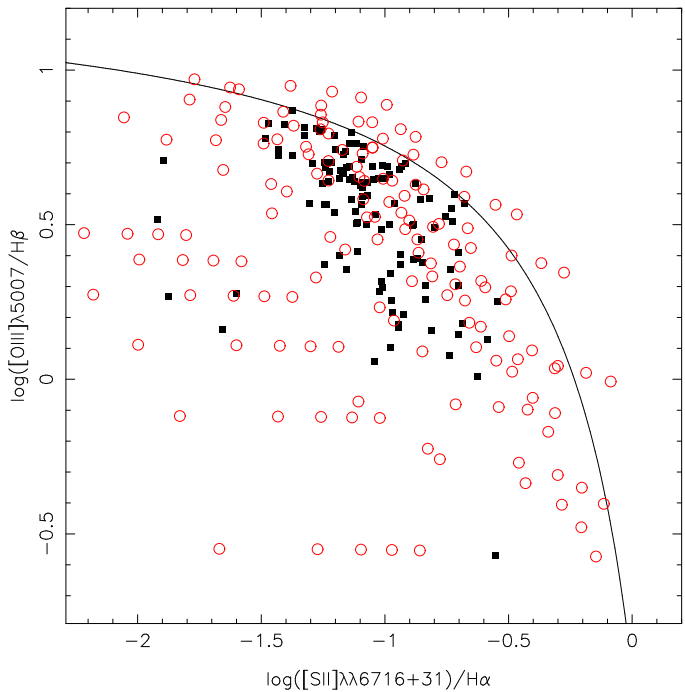}
\caption{$\rm log[O\:III]\lambda5007/H\beta$ vs.\ $\rm log([S\:II]\lambda\lambda6717+31/H\alpha)$ diagnostic diagram.
Solid line, taken from \citet{kewley01}, separates objects ionized by massive stars from those containing active nuclei and/or shock-excited gas. 
Black squares represent the objects in our sample.  Open circles represent  estimations predicted by our models (see Sect.~\ref{mod}).}
\label{fdia}
\end{figure}

Since  \ion{H}{ii} regions and star-forming galaxies are indistinguishable
in diagnostic diagrams (e.g. \citealt{dors13}), these objects were considered jointly in our analysis.
To eliminate objects with a secondary ionizing source, we use the  criterion proposed by \citet{kewley01} to distinguish objects ionized by massive stars from
those containing an active galactic nucleus (AGN) and/or gas shock. Hence all objects with  
\begin{equation}
\rm log[O\:III]\lambda5007/H\beta \: < \: \frac{0.72}{[log([S\:II]\lambda\lambda6717+ 31/H\alpha)]-0.32}
\end{equation}
were selected. In Figure~\ref{fdia} the objects in our sample and a curve representing the  criterion above  are shown.

  In the Appendix,  Table~\ref{tab1} lists the object identification, optical emission-line intensities (relative to H$\beta$=100) and 
bibliographic references of the sample. We obtained optical data of 118 objects.
All emission-line intensities were reddening corrected 
 by the authors of the original works from which we have taken the data. \citet{dors13}   showed that
 effects of using  heterogeneous data sample, such as the one used in this paper, do not yield
any bias on the results of abundance estimations in the gas phase of
star-forming regions.

We also considered emission-line intensities of 143 \ion{H}{ii} galaxies of a sample of 310 galaxies considered by \citet{izotov06a} and selected 
 from the Sloan Digital Sky Survey  \citep{york00} Data Release 3 . We applied a similar selection criterion above but with small changes due to the shorter 
wavelength spectral coverage of the Sloan data (when  [\ion{O}{ii}]$\lambda$3727 is observed [\ion{S}{iii}]$\lambda$9069
 is not, and vice versa, depending on the object redshift). Hence we selected the objects that present the  [\ion{S}{iii}]$\lambda$9069 
 emission lines and [\ion{O}{ii}]$\lambda$7325 instead of [\ion{O}{ii}]$\lambda$3727. 
These objects are also represented in Fig.~\ref{fdia} but are not listed in Table~\ref{tab1}.

Concerning the IR-sample, the  selection criterion was the presence of the flux measurements
 of the emission-lines \ion{H}{i}\,4.05 $\mu$m, [\ion{S}{iv}]\,10.51$\mu$m and [\ion{S}{iii}]18.71$\mu$m.
 We compiled infrared data of 103 objects  classified  as being
  \ion{H}{ii} regions and  nuclei of galaxies containing star-formation regions.
 Only nine objects have both optical and IR data.
   In  the Appendix, in Table~\ref{tab2},  
object identification, 
fluxes of the emission-lines considered and bibliographic references of the sample are listed.
In some cases, indicated in Table~\ref{tab2}, the \ion{H}{i}\,4.05 $\mu$m emission-line fluxes
were computed from  \ion{H}{i}\,12.37 $\mu$m or H\,I\,2.63 $\mu$m fluxes, 
assuming the theoretical ratios  \ion{H}{i}\,4.051$\mu$m/\ion{H}{i}\,12.37$\mu$m=8.2 
and H\,I\,4.051$\mu$m/H\,I\,2.63$\mu$m=1.74 taken from  \citet{storey95}
for $N_{\rm e}=100$ cm$^{-3}$ and $T_{\rm e}=10\:000$ K.

For the objects with emission-line measurements at different spatial positions, indicated in the Table~\ref{tab2},
 the adopted fluxes were the sum (integrated) of the individual ones. 
 The purpose of this procedure is to avoid taking   exclusive emission-lines  from 
outer parts of \ion{H}{ii} regions into account, which the  diffuse gas emission (e.g. \citealt{helmbold05, walterbos98}) 
component can be important but it is not considered in our photoionization models.

  The aperture sizes  in which the optical and infrared data were taken for a same object can be
 different from each other, yielding uncertainties in our results.  In fact, \citet{kewley05} presented
 a detailed analysis of the effect of considering different aperture on determinations
 of physical parameters of galaxies. They have found that systematic and random errors from aperture
effects can arise if fibres capture less than 20 per cent of the galaxy
light. Most of the star-forming regions in our sample can be treated
as point sources, and almost all the object extensions are observed.
Therefore, this effect seems to be negligible for our sample of  objects.

\section{Photoionization models}
\label{mod}
 
 We built a grid of photoionization models using the Cloudy code version 13.03  \citep{ferland13}
 to estimate an  ICF for the sulphur. These models are similar to the ones presented by \citet{dors11}
 and in what follows  the input parameters  are briefly discussed: 
\begin{itemize}
 \item Spectral Energy Distribution ---  The  synthetic spectra  of   stellar clusters with 1 Myr, built with the $STARBURST99$ \citep{leitherer99}  
 assuming the WM-basic stellar atmosphere models by \citet{pauldrach01}, and the 1994 Geneva tracks with standard mass loss  with metallicities 
$Z= 1.0, 0.4, 0.2, 0.05 \: Z_{\odot}$, were considered.

\item Ionization parameter -- The ionization parameter $U$ is defined as $U= Q_{{\rm ion}}/4\pi R^{2}_{\rm in} n  c$, where $ Q_{\rm ion}$  
is the number of hydrogen ionizing photons emitted per second
by the ionizing source, $R_{\rm in}$  is  the distance from the ionization source to the inner surface
of the ionized gas cloud (in cm), $n$ is the  particle
density (in $\rm cm^{-3}$), and $c$ is the speed of light.   
We assumed $R_{\rm in}=4\, \rm pc$, a typical size of a stellar cluster and
 also used by \citet{grazina03} to model a large sample of data of star-forming galaxies. 
The value $n=200\: \rm cm^{-3}$ was assumed in the models,  a typical value
 of \ion{H}{ii} regions located in disks of isolated galaxies (e.g. \citealt{krabbe14}).  
 
We considered the  $\log  Q_{\rm ion}$  
ranging  from  48 to 54 dex, with a step of 1.0 dex.   
From the computed sequence of models for the hypothetical nebulae,
we found $\log U$ ranging  from $\sim-1.5$ to $\sim-4.0$,
typical values of \ion{H}{ii} regions (e.g. \citealt{sanchez15, enrique14, rosa14, priscila14, dors13, bresolin99}).  

\item Metallicity -- The metallicity of the gas phase, $Z$,
was linearly scaled to the solar metal composition \citep{allende01}  
and the values   $Z= 1.0, 0.6, 0.4, 0.2, 0.05 \: Z_{\odot}$ were considered.
 In order to build realistic models, the metallicity of the
nebula was matched with the closest available metallicity of
the stellar atmosphere (see \citealt{dors11} for a discussion about this methodology).
   For the nitrogen, we  computed its abundance  from
the relation between N/O and O/H given by  \citet{vilacostas93}. Although the relation 
between N and O presents a high dispersion (e.g. \citealt{enrique09}) this does
not affect the results of the present study, since we do not use nitrogen emission-lines.
Since  the relation between S/O and metallicity 
is uncertain \citep{enrique06, carolina06},  five grids of models were built with the following values of log(S/O): $-1.31$, $-1.42$ (solar value), $-1.55$
$-1.72$, and $-2.12$.  
\end{itemize}

The presence of internal dust was
considered and the grain abundances  of \citet{hoof01}  were linearly scaled with the oxygen abundance.
The abundances of the  refractory elements Mg, Al, Ca, Fe, Ni and Na were depleted by a factor
of 10, and Si by a factor of 2, relative to the adopted abundances of the gas phase in each model.  The resulting 
geometry was spherical in all models. In total, 175 photoionization models were built. 
 In Fig.~\ref{fdia},  intensities of the line ratios  $\rm log([O\:III]\lambda5007/H\beta)$  and \ $\rm log([S\:II]\lambda\lambda6717+31/H\alpha)$
predicted by the models are also plotted, where it can be seen that the models
 cover  very well the region occupied by the observations.

\section{Determination of ionic abundances}
\label{ionic}

Using the observational data in Table~\ref{tab1}, 
the ionic abundances of $\rm O^{+}$, $\rm O^{2+}$, $\rm S^{+}$ and $\rm S^{2+}$
were computed using direct estimations of the electron temperatures
(following \citealt{dors13}, this method will be called the Visible-lines method).
We also used the observational data in Table~\ref{tab2} to calculate
the  $\rm S^{2+}$ and $\rm S^{3+}$ ionic abundances through  
 infrared emission-lines (this method will be called the IR-lines method).
In what follows, a description of each method is given.

\subsection{Visible-lines method}
\label{visib}

For the objects listed in Table~\ref{tab1}, the electron temperature values and  oxygen 
and sulphur ionic abundances were derived from
 the expressions obtained by \citet{enrique14} and by using 
the same  atomic parameters  used in the version 13.03 of the Cloudy code  and  listed 
in Table~\ref{tab2a}. These parameters  were included in the PyNeb code \citep{luridiana15}
to derive  the oxygen and sulphur abundances as a function of emission-line ratios and electron temperature. 
These equations are valid for the electron temperature range  8000-25000 K and they are presented
in what follows.

For the objects  listed in Table~\ref{tab1},  we calculated the 
electron temperature ($T_{\rm e}$) from the observed line-intensity ratio 
$R_{\rm O3}$= (1.33$\times$$I$[\ion{O}{iii}]$\lambda 5007)$/$I$[\ion{O}{iii}]$\lambda4363$
for the high ionization zone (refereed as $t_{3}$) using the fitted function:
 \begin{equation}
 \label{eqt3}
 t_{3}=0.7840-0.0001357 \times R_{\rm O3}+\frac{48.44}{R_{\rm O3}},
 \end{equation}
with $t$ in units of $10^{4}$K.

 Adopting  the same methodology of \citet{enrique14},
the electron density ($N_{\rm e}$) was computed
from the ratio  $R_{S2}=$[\ion{S}{ii}]$\lambda 6716/\lambda 6731$  and using
the following expression proposed by \citet{hagele08} 

\begin{equation}
\label{eqt3d}
N_{\rm e}= 10^3 \cdot \frac{R_{S2} \cdot a_0(t) + a_1(t)}{R_{S2} \cdot b_0(t) + b_1(t)},
\end{equation}
\noindent with $N_e$  in units of cm$^{-3}$ and $t$ in units of
10$^4$ K.

Using the appropriate fittings and  PyNeb with 
collision strengths   listed in Table~\ref{tab2a}, the coefficients  
of Eq.~\ref{eqt3d} can be written in the form

\[ a_0(t) = 16.054 - 7.79/t - 11.32\cdot t_{2}, \]
\[ a_1(t) = -22.66 + 11.08/t + 16.02\cdot t_{2}, \]
\[ b_0(t) = -21.61 + 11.89/t + 14.59\cdot t_{2}, \]
\begin{equation}
b_1(t) = 9.17 - 5.09/t - 6.18\cdot t_{2}, 
\end{equation}
being $t_{2}$ defined by
 
\begin{eqnarray}
\label{eqo3}
t_{2} =\frac{1.397}{0.385+t_{3}^{-1}}.
\end{eqnarray}
 For the cases where $R_{S2}$ is unresolved, 
a value of $N_{\rm e}=200\: \rm cm^{-3}$ was assumed.

The $\rm O^{2+}$ and $\rm O^{+}$ abundances were computed  
following the relations: 
\begin{eqnarray}
\label{eqt4}
 12+\log \left(\frac{{\rm O^{2+}}}{{\rm H^{+}}}\right) \!\!\!&=&\!\!\! \log \left[ \frac{I(5007)}{I{\rm (H\beta)}}\right]+6.3106  \nonumber\\
                                          &&\!\!\!+\frac{1.2491}{t_{3}}-0.5816 \: \times \: \log  t_{3}   
\end{eqnarray}
and
\begin{eqnarray}
\label{eqt5}
 12+\log \left(\frac{{\rm O^{+}}}{{\rm H^{+}}}\right)  \!\!\!&=&\!\!\! \log  \left[ \frac{I(3727)}{I{\rm (H\beta)}}\right]+5.887 \nonumber\\
                                           &&\!\!\!+\frac{1.641}{t_{2}}-0.543 \times \log t_{2} +10^{-3.94}\:n_{\rm e},
\end{eqnarray}
where $n_{\rm e}=N_{\rm e}/(10^{4}\: \rm cm^{-3})$.

Concerning the SDSS data taken from \citet[][not listed in Table\,\ref{tab1}]{izotov06a},  for the objects with 
redshift $z\: > \:0.02$ in which the [\ion{S}{iii}]$\lambda$9069 was measured, the  [\ion{O}{ii}]$\lambda$3727
is out of the spectral range. Therefore, for this dataset, the $\rm O^{+}$ abundance  was computed using the  
fluxes of the  [\ion{O}{ii}]$\lambda$7320,$\lambda$7330 emission-lines  and the expression also derived using the 
PyNeb code \citep{luridiana15}:

\begin{eqnarray}
 12+\log \left(\frac{{\rm O^{+}}}{{\rm H^{+}}}\right)  \!\!\!&=&\!\!\! \log  \left[ \frac{I(7320+7330)}{I{\rm (H\beta)}}\right]+7.21 \nonumber\\
                                           &&\!\!\!+\frac{2.511}{t_{2}}-0.422 \times \log t_{2} + \\&&10^{-3.40} \: n_{\rm e} ( 1-10^{-3.44}\: \times \: n_{\rm e})\nonumber.
 \end{eqnarray}

For the sulphur ionic abundances, the equations used are: 

\begin{table*}
\caption{Sources of the atomic data   of  sulphur and oxygen 
 ions. }
\vspace{0.3cm}
\label{tab2a}
\begin{tabular}{@{}lcc@{}}
\hline		
                            &\multicolumn{2}{c}{References}   \\
\cline{2-3}
\noalign{\smallskip}		    
Ion                       &      Transition probabilities             &   Collisional strengths       \\
\hline
$\rm S^{+}$          &   \citet{podobedova09}    &         \citet{tayal10} \\
$\rm S^{2+}$        &    \citet{podobedova09}   &         \citet{tayal99}           \\
$\rm S^{3+}$      &     \citet{johnson86}         &         \citet{tayal00}          \\
$\rm O^{+}$          &     \citet{zeippen82}        &       \citet{pradhan06}           \\
$\rm O^{2+}$          &     \citet{storey00}     &       \citet{aggarwal99}            \\
\hline
\end{tabular}
\end{table*}

\begin{eqnarray}
\label{eqt6}
 12+\log \left(\frac{{\rm S^{+}}}{{\rm H^{+}}}\right)  \!\!\!&=&\!\!\! \log  \left[ \frac{I(6717+6731)}{I{\rm (H\beta)}}\right]+5.423 \nonumber\\
                                           &&\!\!\!+\frac{0.941}{t_{2}}-0.37\log t_{2}
\end{eqnarray}
and

\begin{eqnarray}
\label{eqt7}
 12+\log \left(\frac{{\rm S^{2+}}}{{\rm H^{+}}}\right) \!\!\!&=&\!\!\! \log \left[ \frac{I(9069)}{I{\rm (H\beta)}}\right]+6.527  \nonumber\\
                                          &&\!\!\!+\frac{0.661}{t_{S3}}-0.527\log t_{S3}.    
\end{eqnarray}
To derive the $t_{S3}$  temperature for the gas region where the $\rm S^{2+}$ is located,
 we used the  relation  (see \citealt{enrique05})
\begin{eqnarray}
\label{eqo3}
t_{S3}=1.05 \times \: t_{3}-0.08.
\end{eqnarray}

 The electron temperature ($t_{3}$), electron density and ionic abundances calculated 
from the   preceding equations   and using the optical data (Table~\ref{tab1}) are listed
in Table~\ref{tab3} in the Appendix.  Typical errors of  emission-line intensities
are about 10-20  per cent and of   electron temperature determinations  $\sim$500 K, which
yield an uncertainty in ionic abundances of about 0.15 dex (see \citealt{hagele08, kennicutt03, vermeij02a}).
Hereafter, we will assume that the   abundances based on Visible-lines method
have an uncertainty of 0.15 dex.

\subsection{IR-lines method}
\label{corr}

In order to derive  more precise  ionic sulphur abundances, we have taken into account the
 temperature dependence on the emission coefficients   to derive 
$\rm S^{2+}$  and $\rm S^{3+}$ abundances from infrared lines. We computed the $\rm S^{2+}$ 
and $\rm S^{3+}$ ionic fractions from
 [\ion{S}{iii}]\,18.71$\mu$m and  [\ion{S}{iv}]\,10.51$\mu$m emission-lines, respectively,
  and considering the line  $\rm H\:I\: 4.05\: \mu m$  presented in Table~\ref{tab2}. 
  We used the code   PyNeb \citep{luridiana15} and the atomic parameters presented in 
 Table~\ref{tab2a} to derive the equations

\begin{eqnarray}
\label{eqir8}
 12+\log(\frac{{\rm S^{2+}}}{{\rm H^{+}}}) \!\!\!&=&\!\!\! \log \big( \frac{I({\rm 18.71 \mu m})}{I{\rm (H\beta)}}\big)+7.051  \nonumber\\
                                          &&\!\!\!-\frac{0.053}{t_{\rm e}}-0.634\log t_{\rm e} 
\end{eqnarray}
 and 
\begin{eqnarray}
\label{eqir9}
 12+\log(\frac{{\rm S^{3+}}}{{\rm H^{+}}}) \!\!\!&=&\!\!\! \log \big( \frac{I({\rm 10.51 \mu m})}{I{\rm (H\beta)}}\big)+6.218  \nonumber\\
                                          &&\!\!\!+\frac{0.098}{t_{\rm e}}-0.252\log t_{\rm e}.  
\end{eqnarray}

 Since it is not possible to calculate the  electron 
temperature  for most of the objects ($\sim90$\%) in our IR-sample  (presented in Table~\ref{tab2}), 
we  assumed  $T_{\rm e}$=10\,000\,K that implies a certain amount of error.
Variations of $\pm5000$ K in the value of  the electron temperature
in Eqs.~\ref{eqir8} and \ref{eqir9} do the ionic abundance ranges by about $\pm$0.1 dex.
Moreover, for these objects, 
we considered the theoretical relation  $I$(H$\beta$)/$I$(\ion{H}{i}\,4.05 $\mu$m)=12.87
assuming $N_{\rm e}$=100 $\rm cm^{-3}$ and $T_{\rm e}$=10\,000\,K \citep{osterbrock89}.

 Typical uncertainties in IR estimations are 
of the order of 0.1 dex and are caused, mainly,  by the error in the emission-lines \citep{vermeij02a}.
Hereafter, we will assume that the ionic   abundances  calculated from IR-lines method
have an uncertainty of 0.10 dex.


\begin{figure}
\centering
\includegraphics[angle=-90,width=9cm]{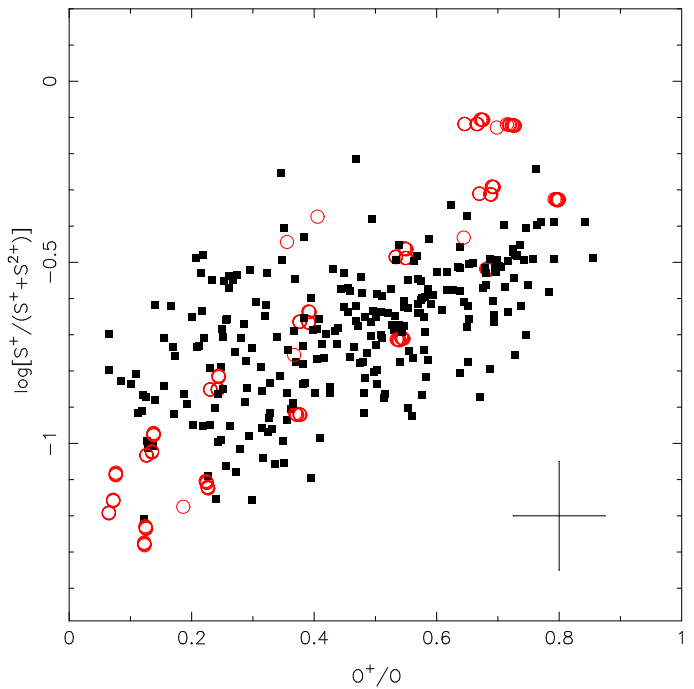}
\caption{Ionic abundances $\rm S^{+}/(S^{+}+S^{2+})$ vs.\ $\rm O^{+}/O$. Black squares represent
observational ionic determinations computed using the  data from Table~\ref{tab1} and the Visible-lines method
(see Sect.~\ref{corr}). Open circles represent ionic abundances predicted by our models (see Sect.~\ref{mod}).
 The error bar represents  typical uncertainties  as defined in Sect~\ref{visib}.}
\label{f0}
\end{figure}

\begin{figure}
\centering
\includegraphics[angle=-90,width=9cm]{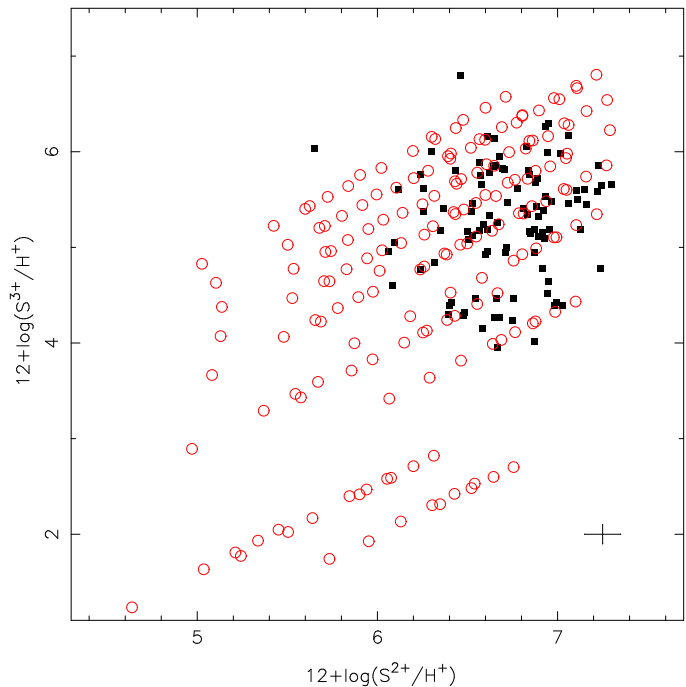}
\caption{Such as Fig.~\ref{f0} but for the
Ionic abundances $\rm S^{3+}/H^{+}$ vs.\ $\rm S^{2+}/H^{+}$ computed using the IR-sample {\rm (Table~\ref{tab2})}
and the IR-lines method (see Sect.~\ref{corr}). }
\label{f1}
\end{figure}

\section{Results}
\label{result}

\subsection{Theoretical-ICF}
\label{ifs1}

 We  derived a theoretical ICF for the sulphur based on 
the photoionization model results described in Sect.~\ref{mod}.
To verify how representative are our models of real \ion{H}{ii} regions,
in Fig.~\ref{f0}, the ionic abundance ratio  $\rm S^{+}/(S^{+}+S^{2+})$ against
the ionization degree $\rm O^{+}/O$ calculated from the  data from Table~\ref{tab1} 
and using the Visible-lines method  are compared with   those predicted by the models.
 The theoretical ionic values considered are the ones weighted over the volume of the hypothetical nebulae.
We can see that the models occupy  the most part of the region where   the observational data are located
and they reproduce the tendency of $\rm S^{+}/(S^{+}+S^{2+})$ increases with $\rm O^{+}/O$.
However,  there is a region  occupied by observational  data with  $[\rm S^{+}/(S^{+}+S^{2+})]\: \ga \: -1$ and  $\rm (O^{+}/O) \: \ga \: 0.2$
 not covered by the models.  This seems to be  not crucial for the present analysis 
 since similar ICFs can be derived from both models and
 observations, as we are presenting in this paper.  
 
 In Fig.~\ref{f1},  the  $\rm S^{3+}/H^{+}$  and \ $\rm S^{2+}/H^{+}$
 abundances calculated using the IR-lines method and the IR-sample 
and those predicted by the models are shown. Again, we can see that the models cover
the region occupied by the  observations.

 \begin{figure}
\centering
\includegraphics[angle=-90,width=9cm]{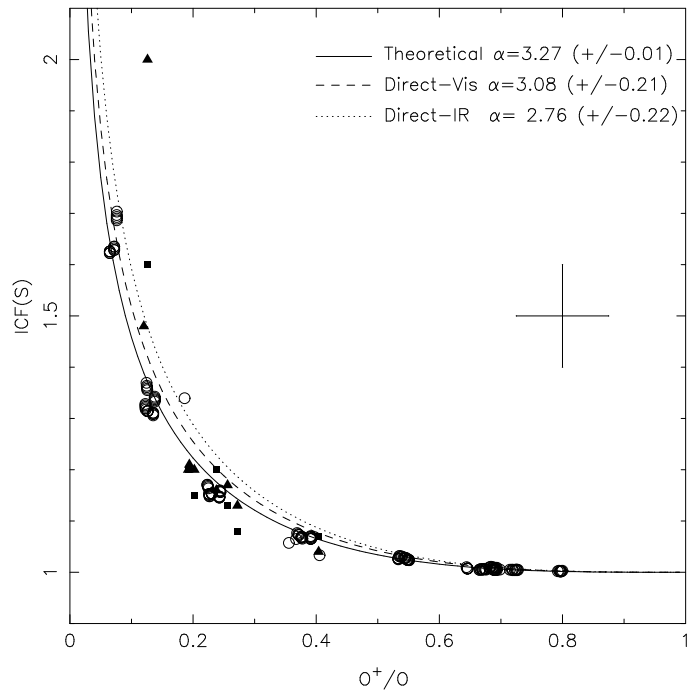}
\caption{Ionization Correction Factor for the sulphur vs.\  $\rm O^{+}/O$.  
 Squares and triangles  represent  direct estimations (see Table~\ref{tab3})
of the ICF taking into account the $\rm S^{2+}$ ionic abundance values estimated from  
the Visible-sample (Direc-Vis) and the IR-sample (Direct-IR), respectively.
Circles represent estimations of our models.
Curves represent the fittings to the Eq.\ \ref{eq0}: Solid line
shows the best fit obtained using our models and  dashed and dotted lines the ones obtained using the 
observational estimations,  as indicated. The $\alpha$ values of the best fits are indicated in the legend.
The error bars represent  typical uncertainties  as defined in Sect~\ref{corr}.}
\label{f4}
\end{figure} 

The predictions of the  models were used to
 compute an ICF for the sulphur  defined by:
 \begin{equation}
 \label{eqicf}
{\rm ICF(S^{+}+S^{2+})}=\frac{{\rm S/H}}{{\rm (S^{+}+S^{2+})/H^{+}}},
\end{equation} 
where  $\rm S/H$ is the ratio between the total sulphur  and the 
hydrogen abundances. Assuming the expression suggested by \citet{stasinska78a} and presented in 
Eq.\ \ref{eq0},  we found $\alpha=3.27\pm0.01$ from a fitting to our model results.

\subsection{Direct ICFs}
\label{dir}

When emission-lines of the main ionization stages  of an element are observed,
 it is possible to calculate the total abundance of the element and thus, derive an ICF. 
 Therefore,  following the methodology presented by \citet{vermeij02a} and  \citet{enrique06},
we used the Visible and IR samples and the 
equations presented in Section~\ref{ionic} to derive direct values for the  sulphur ICF for the common objects
in both samples assuming
\begin{equation}
 \label{eqicf}
{\rm ICF(S^{+}+S^{2+})}=\frac{{\rm S^{+}+S^{2+}+S^{3+}}}{{\rm S^{+}+S^{2+}}}.  
\end{equation}

This  was possible only for nine objects. The $\rm S^{2+}$ can be estimated using the Visible data  and/or using the IR data.
Hence, for each object, we have two independent estimations of  its sulphur ICF,
these two values are named Direct-Vis and Direct-IR ICFs.

The identification of the nine objects for which was possible to compute the ICF by the procedure described above,
the electron temperature ($t_{S3}$) and the ionic abundance va\-lues are listed
in Table~\ref{tab3}, while the $\rm O^{+}/O$ ratio and the ICF values are
presented in Table~\ref{tab3a}.
For Hubble\,V  and I\,Zw\,36 were only possible to compute the $\rm S^{2+}$ ionic abundance 
via the IR-method  because the [\ion{S}{iii}]$\lambda$9069,$\lambda$9532 emission-lines are not available in the literature.
These are the only two objects in the subsample that do not fulfil the selection criterion to be in the Visible-sample but 
were included here because they contribute to a better estimation of the Direct-IR sulphur ICF.
 The  difference in the $\rm S^{2+}$  abundances calculated from Visible and IR lines methods
has an average value of $\sim$0.15 dex, with the maximum value of $\sim$0.35 dex. 
 In the subsequent paper of this series, we will use  photoionization models with
abundance variations along the radius of the hypothetical nebula in order to
investigate the source of this discrepancy.\footnote{Similar analysis but applied 
for neon ionic abundances can be found in  \citet{dors11}.}

In Fig.~\ref{f4} the direct sulphur ICF values as a function of $\rm O^{+}/O$ are plotted together with 
the corresponding fittings. We found  $\alpha=2.76 \pm0.22$
when $\rm S^{2+}$ is computed by the IR-method and $\alpha=3.08 \pm0.21$ when 
the Visible-method is considered.   We can note in Fig.~\ref{f4} that the two fits for
the estimations based on IR and Visible methods (red and blue lines) seem to be  not satisfactory
for $\rm O^{+}/O \la 0.2$, i.e. for the regime of high excitation.
Similar result was  found by \citet{vermeij02a}.
A larger number of direct ICF estimations for objects with  high excitation   is clearly need to improve
the results for this regime.

The error in the  Direct-ICF value is  due to the  uncertainties  of ionic  abundance determinations ($\rm S^{+}, S^{2+}, O^{+}$, and $\rm O^{2+}$) 
and   due to the discrepancy  between the  abundance  of $\rm S^{2+}$  calculated  via
Visible and IR methods \citep{dors13, vermeij02a}.
Based on the results of  \citet{vermeij02a}, we assumed an average error
 of 0.2 for the Direct-ICF and 0.15 for  $\rm O^{+}/O$, obtained from ionic estimations of \citet{kennicutt03}.
  These uncertainties  yield  an  error in the  total sulphur abundance  of only  $\sim$10\%.

\begin{table*}
\caption{Electron temperatures ($t_{S3}$) and sulphur ionic abundances estimated for the Visible and IR samples.}
\vspace{0.3cm}
\label{tab3}
\begin{tabular}{@{}lccccc@{}}
\hline  	
\noalign{\smallskip}
Object                         &          $t_{\rm e} (10^{4}$K)        &  $\log(\rm S^{+}/H^{+})_{Vis}$&  $\log(\rm S^{2+}/H^{+})_{Vis}$     &   $\log(\rm S^{2+}/H^{+})_{IR}$         & $\log(\rm S^{3+}/H^{+})_{IR}$     \\[3pt]
N160A1                       &           0.92                                  &         -6.24                             &             -5.20                                &          -5.31                                     &           -6.03                                \\
N160A2                      &           0.88                                  &         -6.24                             &             -5.18                                &          -5.39                                      &           -6.22                                \\
N4A                           &            0.94                                  &         -6.41                              &            -5.27                                 &         -5.15                                      &           -5.93                                 \\
N66                           &            1.18                                 &         -6.53                              &            -5.69                                  &         -5.72                                      &           -6.35                                 \\
N157-B                      &            1.29                                 &        -6.09                               &            -5.49                                  &         -5.30                                      &           -6.57                                \\
N88-A                        &            1.41                                 &       -6.87                               &            -6.05                                   &         -6.40                                      &          -6.28                                  \\    
N81                            &            1.26                                 &       -6.72                                &            -5.81                                  &         -6.00                                     &          -6.62                                   \\
Hubble\,V$^{\rm a}$   &             1.09                                &        -6.68                                &           ---                                       &        -5.58                                      &          -6.21                                    \\
I\,Zw\,36$^{\rm a}$      &             1.61                               &       -6.90                                 &           ---                                        &       -5.81                                      &          -6.08                                    \\
\hline
\end{tabular}
\begin{minipage}[c]{2\columnwidth}
$^{\rm a}$See text for an explanation about the inclusion of these two particular objects.
 \end{minipage}
\end{table*}

\begin{table*}
\caption{$\rm O^{+}/O$ ionic abundances and direct sulphur ICFs estimations using the Visible-lines and the IR-lines methods.}
\vspace{0.3cm}
\label{tab3a}
\begin{tabular}{@{}lccc@{}}
\hline                     
      Object             &     $\rm O^{+}/O$                          &\multicolumn{2}{c}{ICF}   \\ 
\cline{3-4}			     	
\noalign{\smallskip}
             &                                    &    Vis    & IR        \\
N160A1       &      0.256               &    1.13   &   1.17    \\
N160A2       &      0.272               &    1.08   &   1.13    \\
N4A             &     0.238               &    1.20    &  1.16    \\
N66             &      0.192               &    1.20   &  1.20     \\
N157-B       &       0.404              &    1.07   &  1.04     \\
N88-A         &       0.126               &    1.60   &  1.98     \\
N81            &       0.202               &    1.15   &   1.20     \\
Hubble\,V    &      0.194                &    ---     &   1.21      \\
I\,Zw\,36      &      0.120                &   ---      &  1.49      \\
\hline
\end{tabular}
\end{table*}

\section{Discussion}
\label{disc}

In their pioneer paper, \citet{peimbert69}  obtained   photoelectric observations of the \ion{H}{ii}
regions Orion, M\,8 and M\,17 and suggested that the total abundance of the sulphur can be obtained
by using an ICF defined by the ionic abundance ratio ($\rm O^{+}+O^{2+}$)/$\rm O^{+}$. 
 This empirical approach is based on the similarity between the ionization potentials
of   $\rm S^{2+}$ and   $\rm O^{+}$. During the next decades,  sulphur ICFs had been mainly derived from
the analytical expression suggested by \citet{stasinska78a}, and the  $\alpha$ value
of this original prescription have been largely discussed.
For example, data obtained  with the
Infrared Space Observatory by \citet{vermeij02} became, possibly, the first test for the
$\alpha$ value, since direct estimations of sulphur ICFs were possible. These authors
showed that an $\alpha$ value equal to 2, as suggested by \citet{french81}, overpredicts
the $\rm S^{3+}$  ionic abundance, in concordance with the result previously obtained by \citet{garnett89}. 
From their observational data, \citet{vermeij02a} concluded that $\alpha=3$ is a more reliable 
value, at least for $\rm O^{+}/O > 0.2$.

 Despite  ICFs could be obtained  from  direct calculation of ionic abundances \citep{vermeij02a}
or even from ionization potential considerations  (e.g. \citealt{peimbert69, french81}), 
ICFs based on grids of photoionization mo\-dels of nebulae are more reliable
because all ionization stages of a given ion as well as several physical processes
(e.g. charge transfer reactions) are taken into account in the calculations \citep{stasinska02}. 
In the present work,  we built a grid of photoionization models assuming a
large range of nebular parameters (e.g. $Z$, $U$, S/O) and derived a  theoretical sulphur ICF.
Based on the agreement between the model predictions and data of a large sample
of objects, 
we suggested  an $\alpha$ value of $3.27 \pm0.01$ in the Stasi\'nska formulae.
This value is  somewhat higher than the one derived
by \citet{vermeij02a}, but it is in consonance with the one  
derived through direct ionic estimations ($\alpha=3.08 \pm0.21$) based mainly on the Visible-line method (Direct-Vis ICF).

\begin{figure*} 
\centering
\includegraphics[angle=-90,width=1.0\columnwidth]{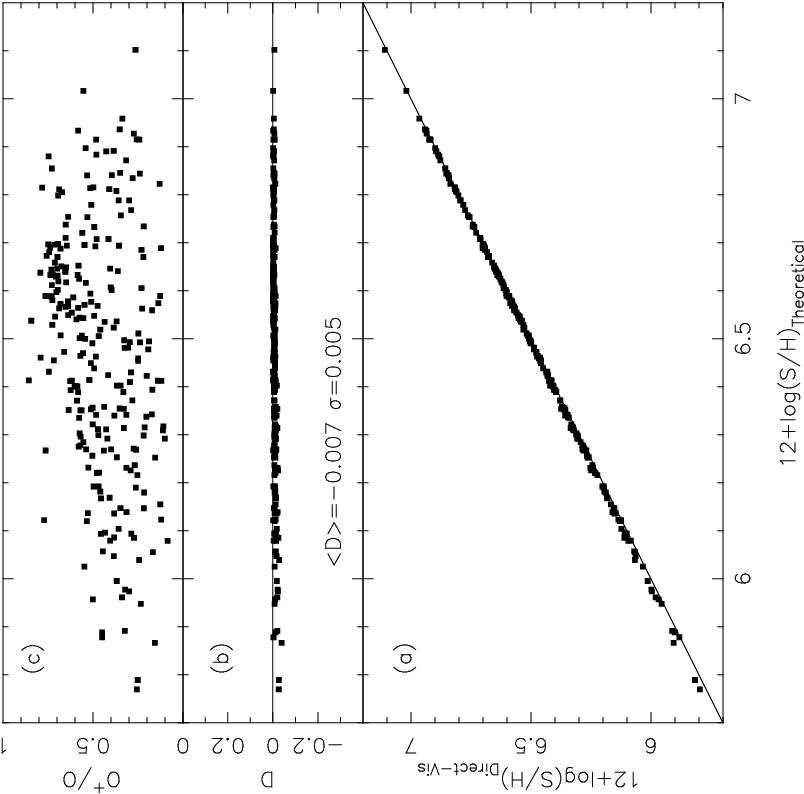}\hspace{0.4cm}
\includegraphics[angle=-90,width=1.0\columnwidth]{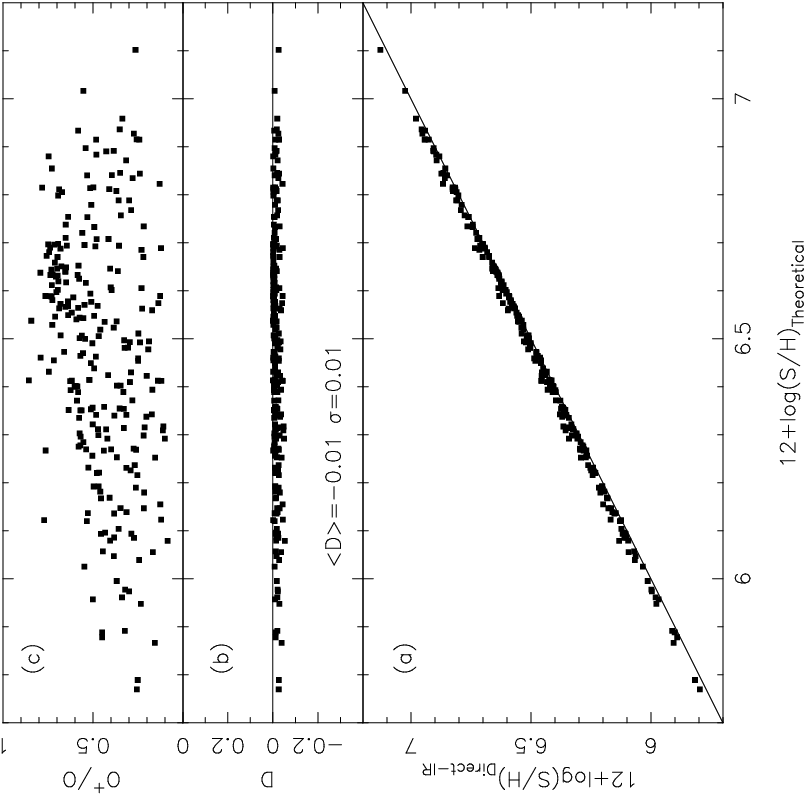}\\\hspace{0.4cm}
\caption{Panel \textit{a}: comparison between the S/H total abundances 
obtained for the objects in the Visible-sample 
applying the  Theoretical ICF and,  Direct-Vis  (left plot) and
Direct-IR (right plot) ICFs, as indicated.
Panel \textit{b}:  differences between the estimations using the considered ICFs with they   
average value ($\rm <D>$) and its dispersion ($\sigma$) indicated. 
Panel \textit{c}: $\rm O^{+}/O$ ratio for each estimation.} 
\label{f5}
\end{figure*}

With the aim to compare the S/H total abundances yielded by the
use of different ICFs, we considered the relation:  
 \begin{equation}
 \rm \frac{S}{H}= ICF(S) \: \times \: \frac{S^{+}+S^{2+}}{H^{+}}, 
 \end{equation}
using the $\rm S^{+}$ and $\rm S^{2+}$ ionic abundances estimated for the objects 
in our Visible-sample via the Visible-lines method.
 Firstly, we compared the S/H abundances derived through the  Theoretical ICF ($\alpha=3.27\pm0.01$),  
with those derived using the Direct-Vis ICF ($\alpha=3.08\pm0.21$) and the Direct-IR ICF ($\alpha=2.76\pm0.22$). 
In   panels \textit{a}  of Fig.~\ref{f5}  these comparisons 
are shown. In this figure we also plotted the differences (D) between the S/H total abundances estimations (panels \textit{b})
and the $\rm O^{+}/O$ ratio (panels \textit{c}). 
It can be seen that the   Theoretical ICF yields S/H 
total abundances in excellent agreement with those given by the Direct-Vis and Direct-IR ICFs,
 with an average difference $\rm <D> \approx 0.00$ and  dispersions of  0.005 dex and 0.01 dex respectively,
  independently  of the ionization degree that is sampled by the $\rm O^{+}/O$ ratio.   
  
\begin{figure*} 
\centering
\includegraphics[angle=-90,width=0.9\columnwidth]{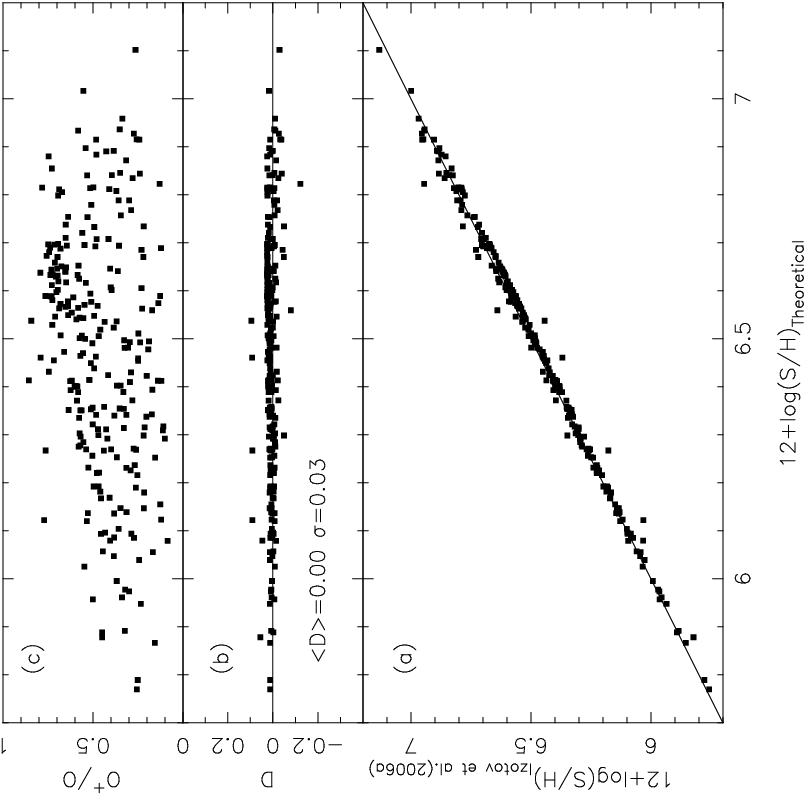}\hspace{0.4cm}
\includegraphics[angle=-90,width=0.9\columnwidth]{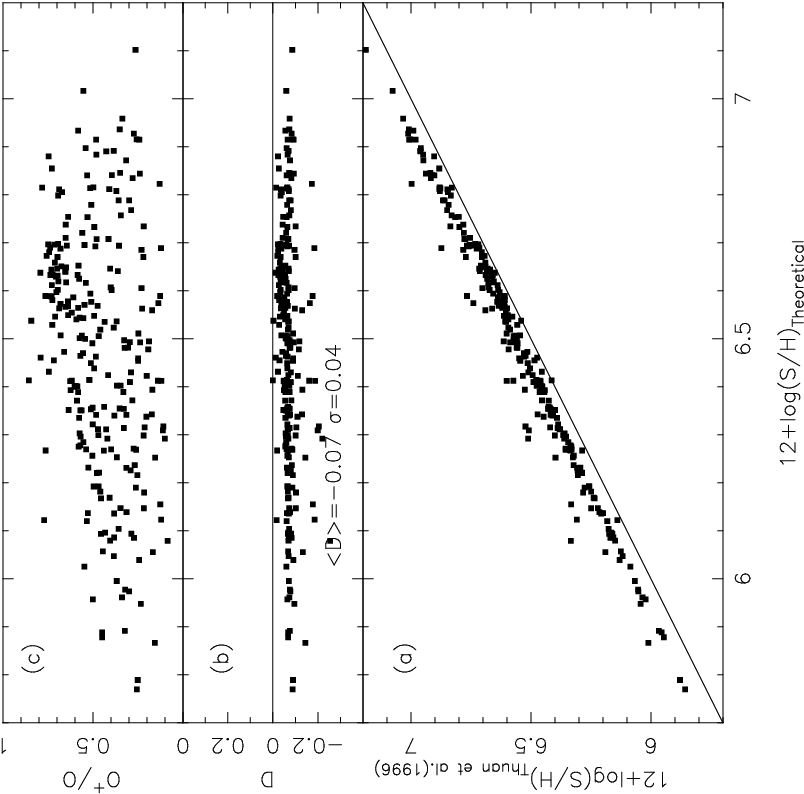}\\\vspace{0.4cm}
\includegraphics[angle=-90,width=0.9\columnwidth]{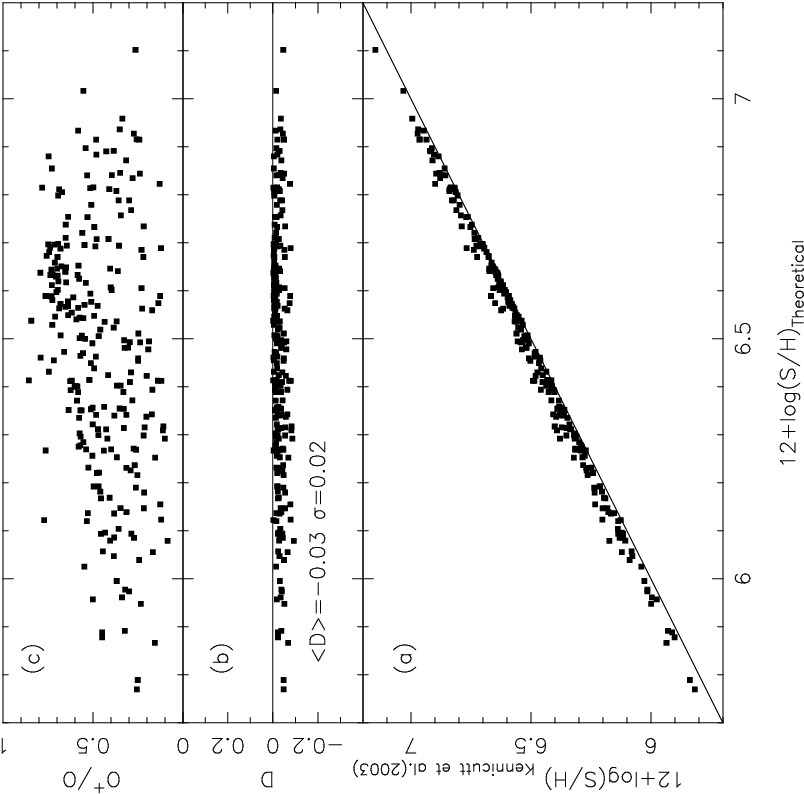}\hspace{0.4cm}
\includegraphics[angle=-90,width=0.9\columnwidth]{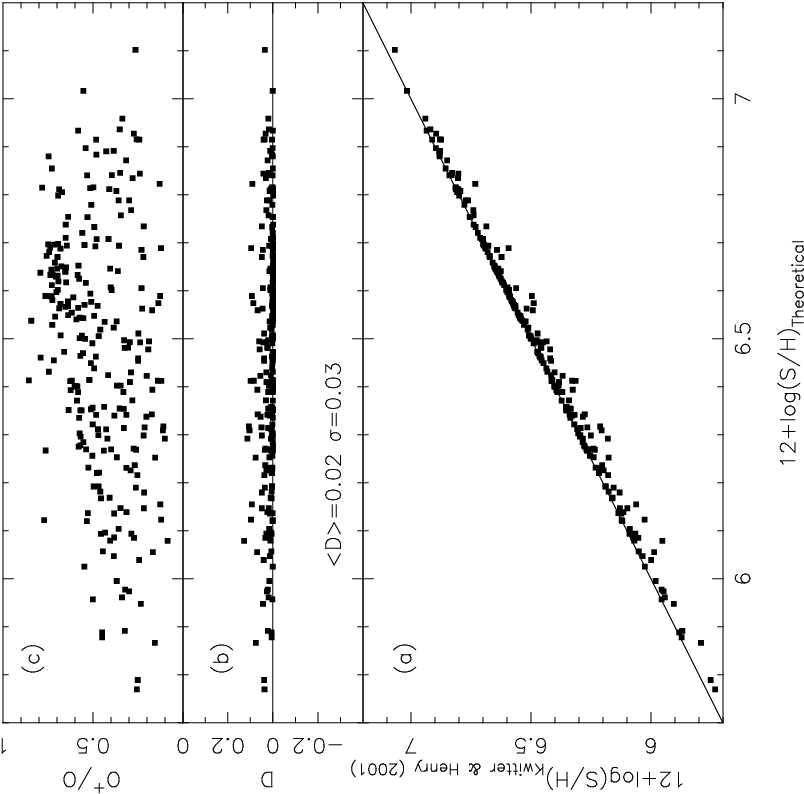}
\caption{Idem Fig.~\ref{f5} for different ICFs from the literature, as indicated.} 
\label{f5a}
\end{figure*}

\begin{figure*}
\centering
\includegraphics[angle=-90,width=9cm]{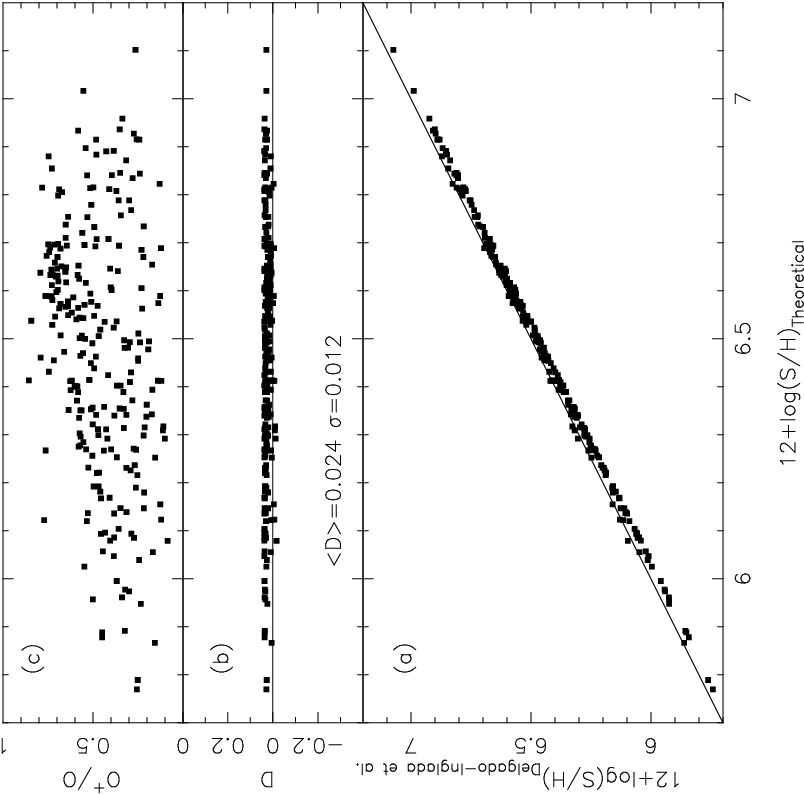}
\caption{Idem Fig.~\ref{f5} for the ICF proposed by \citet{gloria14}}
\label{gloria}
\end{figure*}

Secondly, we also compare the S/H total abundance estimations based on our  Theoretical ICF  with the ones obtained
 using some ICFs proposed
in the literature. In what follows a brief description of these ICFs is presented. 

\begin{itemize}
\item  Kennicutt et al.\ ICF-- 
\citet{kennicutt03}   proposed  to use  $\alpha=2.5$   for typical \ion{H}{ii} regions.
This is an average value obtained from the photoionization models grid calculated by \citet{garnett89}.
The same $\alpha$ value was obtained by \citet{enrique06} from optical and IR data.

\item Izotov et al.\ ICF--   \citet{izotov06a}  used a
grid of  photoionization models by \citet{grazina03} built assuming spectral energy distributions calculated with the
 $Starburst99$ \citep{leitherer99}  and  stellar atmosphere models by \citet{smith02}
 to derive an expression for the sulphur ICF. These authors  derived  
ICFs considering three metallicity  regimes: low [$\rm 12+\log(O/H) \: < \: 7.6$], intermediate 
[$\rm 7.6 \: < \: 12+log(O/H) \:< \:8.2$] and high [$\rm 12+log(O/H)\: > \: 8.2$], which are given by:
\begin{eqnarray}
{\rm ICF}({\rm S })\!\!\!\!&=&\!\!\!\! 0.121x+0.511+0.161/x, \ {\rm low\ }Z, \nonumber\\
                   \!\!\!\!&=&\!\!\!\! 0.155x+0.849+0.062/x, \ {\rm inter.\ }Z, \nonumber\\
                   \!\!\!\!&=&\!\!\!\! 0.178x+0.610+0.153/x\,\,\,  \ {\rm high\ }Z, \nonumber
\end{eqnarray} 
where $x=\rm O^{+}/O$.

\item Thuan et al. ICF-- \citet{thuan95}, who used the results of photoionization models grid built
by  \citet{stasinska90} and NLTE atmosphere models by \citet{mihalas72},
derived
\[
{\rm ICF} \!=\! \Big[0.013+x \, \big[5.10+x \, \big(-12.78+  x \, (14.77-6.11 \, x) \big) \big] \Big]^{-1}.
\]

\item Kwitter \& Henry ICF-- \citet{kwitter01}  built a grid of photoionization models 
considering a blackbody as the ionizing source  in order to derive  sulphur  ICFs 
for planetary nebulae  that, in principle, it can be employed for \ion{H}{ii} regions. 
These authors proposed
\[
{\rm ICF(S)}=e^{-0.017+(0.18 \, \beta) -(0.11 \, \beta^2) +(0.072
\, \beta^3)}, 
\]
where $\beta=\rm \log(O/O^{+})$.  
\item Delgado-Inglada et al. ICF ---  \citet{gloria14} computed a large grid of photoionization models 
in order to derive new formulae for ICFs of several elements  to be applied in  studies of planetary nebulae.  The expression derived 
by these authors  to calculate the  total abundance S/H can be to write in the form

 \[
\rm  \frac{S}{H}=ICF(S) \: \times \: \frac{S^{+}+S^{2+}}{O^{+}}  \: \times \: \frac{O}{H}, 
\]
where
\[
{\rm ICF(S)}=\frac{-0.02-0.03w-2.31w^{2}+2.19w^{3}}{0.69+2.09w-2.69w^{2}} , 
\]
and
\[
w=\rm O^{2+}/O.
\]

\end{itemize}
 
In  Fig.~\ref{f5a} (panels \textit{a}) a comparison between S/H 
total abundance estimations based on our  Theoretical ICF and  
those from the literature are shown.
In this figure we also show  the difference (D) between these estimations 
(panels \textit{b}) and the $\rm O^{+}/O$ ratio (panels \textit{c}). 
 Taking into account the typical errors found in the S/H total abundance estimations \cite[see e.g.][]{hagele08}
 and the dispersion ($\sigma$) of the average differences ($\rm <\!D\!>$), it might seem that the different S/H 
 estimations are in agreement. 
 However, with exception of the  ICF of  \citet{izotov06a}, 
there are clear systematic differences between the values derived through the use of our  Theoretical
 ICF and from the other   ICFs.
Moreover,  difference in S/H abundances obtained from  distinct ICFs  can be not negligible
when  only an individual object is considered. In fact,  
we noted that it could reach up to about  0.3 dex for the low metallicity regime (see Fig.~\ref{f5}).

\begin{figure}
\centering
\includegraphics[angle=-90,width=9cm]{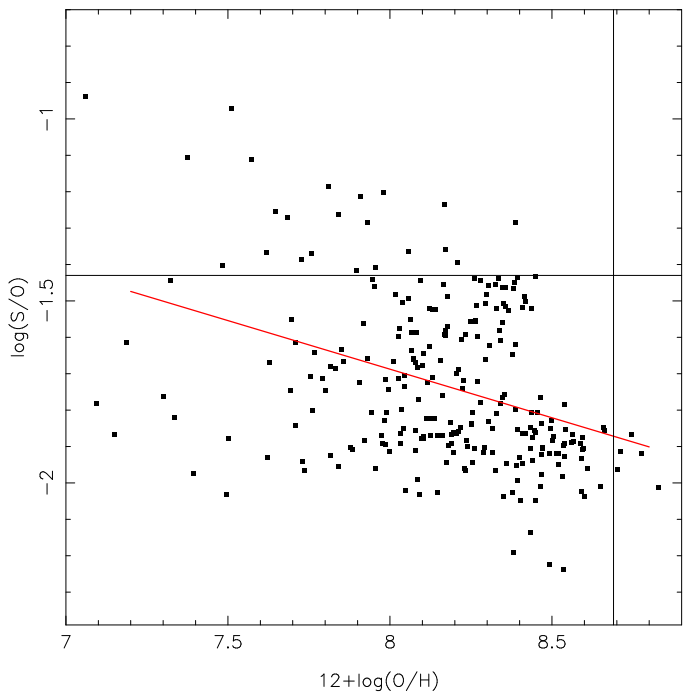}
\caption{Relation between log(S/O) and 12+log(O/H) ratios using the Visible-lines method and  our Theoretical ICF.
 Black lines represent the  solar S/O and O/H abundance ratios  derived using the oxygen abundance from  \citet{allende01} and
 sulphur abundance from  \citet{grevesse98}. Red lines represent linear regressions to the observational estimations. 
  Coefficients of this regression are given in Table~\ref{tab4}.}
\label{f6}
\end{figure}

Concerning  the  ratio between sulphur and oxygen abundances,    
 several studies have addressed the investigation about the variation of  S/O  with O/H
 in individual galaxies (e.g. \citealt{croxall15, berg13, skillman13, lopez09, kennicutt03, vermeij02a, garnett97, christensen97, vilchez88})
or  in a general context (e.g. \citealt{guseva11, hagele08, hagele12, enrique06, carolina06, henry99, izotov97}).
Most of these results indicates that the ratio S/O  appears to be  constant 
with the metallicity, which argues  that either these elements are produced by massive stars within  a similar mass range 
or by stars with  a distinct mass interval but  being formed with an universal IMF
\citep{henry99}. However, when a large sample of data is considered,  the dispersion 
found is very large and the assumption of a constant S/O ratio is 
questionable  \citep{hagele08, hagele12, enrique06, carolina06}. 
Therefore, with the goal of studying the relation of the S/O ratio with the metallicity 
(traced by the O/H abundance),  we used the data listed in Table~\ref{tab1}
and all the ICFs considered in the present work to calculate S/O and O/H ratios via the Visible-lines 
method. The Direct-Vis ICF was not considered  since its $\alpha$ value is very similar to that of the  Theoretical one. 
 In Fig.~\ref{f6} only the  estimations obtained from the Theoretical ICF is shown.
For estimations from other ICFs (not shown), similar results were obtained.
The solar values $\rm \log(S/O)_{\odot}= -1.43$  and $\rm 12+\log(O/H)_{\odot}= 8.69$
derived using the sulphur abundance from 
\citet{grevesse98} and the oxygen one from \citet{allende01} are also indicated.
 We can see in this figure that most of the objects 
present  subsolar S/O  and O/H abundance ratios.
 Interestingly, for the  extreme low metallicity regime, some of the objects reach 
very high  S/O abundance ratios. Since the dispersion is high and the number of objects is much lower than for the high 
metallicty regime, more data are needed to confirm this result.

We  also performed
a fit to these data, assuming a linear regression   without taking into account the individual
errors. In  Table~\ref{tab4},   the coefficients of the fittings,
and the linear regressions considering  all ICFs are listed.   
We found that the S/O ratio decreases with metallicity, yielding a mean
slope of about $-0.27$   with all the fitted slopes in agreement within the estimated errors.
We also obtained  the average values for log(S/O) estimated via the different ICFs and considering the three different metallicity
 regimes. These values and the number of 
 objects used to calculate them are also listed in Table~\ref{tab4}.
Considering all the metallicity regimes together and all the considered ICFs we found an average 
$<\log(\rm S/O)> = -1.72 \pm0.03$.
Despite the dispersion, when  low, intermediate and high metallicities regimes are 
separately considered, we note a decrease in S/O   when the metallicity increases.
For low and high metallicity regimes we derived  mean values of
 $<\log(\rm S/O)>$  $-1.53\pm 0.05$  
  and  $-1.78 \pm 0.02$, respectively.  Similar results were also derived by \citet{diaz91}, \citet{vilchez88} 
  for M51 and M33 galaxies and by  \citet{shaver83} for  Milk Way.
  
\begin{table*}
\caption{Coefficients of the linear regressions $\rm \log(S/O) = \: a \times \:  \big[12+\log(O/H)\big] + b$ fitted to
the data plotted in Fig.~\ref{f6} taking into account each of the considered ICF. Mean values for the abundance ratio $\rm \log(S/O)$
considering all the metallicity range as well as the corresponding ones for the low [$\rm 12+\log(O/H) \: < \: 7.6$], intermediate 
[$\rm 7.6 \: < \: 12+log(O/H) \:< \:8.2$] and high [$\rm 12+log(O/H)\: > \: 8.2$] metallicity regimes, and the number of objects 
used in these calculations are shown.}
\vspace{0.3cm}
\label{tab4}
\begin{tabular}{@{}lccccccc@{}}
\hline	
                              &    &  & 	 \multicolumn{4}{c}{$<\log(\rm S/O)>$} \\
			      &	   &   & \multicolumn{4}{c}{Metallicity regime} \\
\noalign{\smallskip}				    
\cline{4-7}				    
\noalign{\smallskip}
                             & a  &b  &  All                                  & Low                                &  Interm.                                &   High    \\
Number objects      &     &  & 	261			       & 14                                  &    114                                     &   133        \\

\hline
 ICF                       &  &  &                                        &                                        &                                             &              \\ 
\hline
\noalign{\smallskip} 	
Direct-IF                & $-0.28\pm0.04$   &  0.60$\pm0.33$     & $-1.71\pm0.22$                 &   $-1.52\pm0.36$             &   $-1.66\pm0.20$                 & $-1.78\pm0.20$    \\
Theoretical            &   $-0.26\pm0.04$  &  0.45$\pm0.33$    & $-1.73\pm0.22$                &   $-1.55\pm0.38$             &   $-1.68\pm0.20$                 & $-1.79\pm0.19$     \\
Kennicut et al.        &   $-0.30\pm0.04$  &   0.72$\pm0.33$   & $-1.70\pm0.22$                 &  $-1.50\pm0.35$             &   $-1.64\pm0.20$                 & $-1.77\pm0.20$     \\
Izotov et al.            &   $-0.28\pm0.04$  &   0.60$\pm0.33$   & $-1.73\pm0.22$                 &   $-1.53\pm0.31$              &  $-1.68\pm0.20$                 & $-1.80\pm0.20$    \\
Thuan et al.            &  $-0.32\pm0.04$   &  0.94$\pm0.33$    & $-1.66\pm0.23$                  &   $-1.44\pm0.32$             &   $-1.59\pm0.20$                 & $-1.73\pm0.20$     \\
Kwitter \& Henry      &  $-0.23\pm0.04$  &   0.13$\pm0.34$   &  $-1.75\pm0.22$                 &   $-1.59\pm0.41$             &   $-1.71\pm0.21$                  & $-1.79\pm0.19$     \\
Delgado-Inglada et al. & $-0.26\pm0.04$ &  0.41$\pm0.33$   &  $-1.75\pm0.22$                 &   $-1.56\pm0.38$             &   $-1.71\pm0.21$                 &  $-1.81\pm0.19$                            \\
 \hline
\end{tabular}
\end{table*}


 \section{Conclusions}
 \label{conc}
 We  built a grid of photoionization models combined  with 
stellar population synthesis models  to derive  Ionization Correction Factors (ICFs) for the sulphur.
The reliability of these ICFs was  obtained from the agreement between  ionic abundances
predict by the models and those calculated  through  optical and infrared spectroscopic data of star-forming regions with a very wide range in metallicity 
($\rm 7.0 \: \la \: 12+\log(O/H) \: \la 8.8$) and ionization degree    ($\rm 0.1\: \la \:  O^{+}/O  \: \la 0.9$).
From our results, we suggest   $\alpha=3.27 \pm 0.01$  to be  used in the classical Stasi\'nska formula. This $\alpha$ value 
is in consonance with the one derived from direct estimations based on spectroscopic data of a small sample of objects.
A comparison of the S/H total abundance derived by us for the objects in our visible sample and considering different ICFs proposed in the literature was performed.
Although, in average, the differences between these determinations are similar to the  uncertainties in the S/H estimations, we noted that it could reach up to about  0.3 dex for the low metallicity regime.
 Finally, the highest S/O abundance ratios are derived for objects
with extreme low metallicity values.
Indeed, a tendency of the S/O ratio to decrease with the metallicity was found, independently of the considered ICF.

\section*{Acknowledgments}
We are very grateful to the anonymous referee for his/her complete and
deep revision of our manuscript, and very useful comments and
suggestions that helped us to substantially clarify and improve our work
and  to Dr. Gary Ferland for making available
the Cloudy programme.

\section{Appendix} 

\begin{table*}
\caption{Dereddened line fluxes (relative to H$\beta$=100.0) compiled from the literature.
 The intensity of the line [\ion{O}{ii}]$\lambda$3727 represents the sum of the
lines [\ion{O}{ii}]$\lambda$$\lambda$3726+ 29}
\vspace{0.3cm}
\label{tab1}
\begin{tabular}{@{}lcccccccc@{}}
\hline		 
\noalign{\smallskip}  
 Object  & [\ion{O}{ii}]$\lambda$3727 & [\ion{O}{iii}]$\lambda$4363 & [\ion{O}{iii}]$\lambda$5007 & [\ion{S}{ii}]$\lambda$6717 & [\ion{S}{ii}]$\lambda$6731 & [\ion{S}{ii}]$\lambda$6725$^{b}$ & [\ion{S}{iii}]$\lambda$9069  & Ref. \\
\noalign{\smallskip}
H1105              &   185.0   &   1.4   &    316.0     &    13.1     &    11.0     &      ---     &    36.2$^{a}$        &  1 \\
H1159              &   198.0   &   1.9   &    317.0     &    17.6     &    12.2     &      ---     &    27.7$^{a}$        &  1 \\   
H1170              &   308.0   &   1.6   &    201.0     &    33.4     &    23.3     &      ---     &   48.6$^{a}$        &  1 \\
H1176              &   160.0   &   2.4   &    369.0     &    13.4     &     9.6     &      ---     &    32.4$^{a}$        &  1 \\
H1216              &   151.0   &   4.7   &    473.0     &    11.0     &     7.9     &      ---     &    23.7$^{a}$        &  1 \\
H128               &   145.0   &   1.7   &    391.0     &    13.3     &    10.0     &      ---     &    29.8$^{a}$        &  1 \\
H143               &   219.0   &   2.3   &    284.0     &    23.2     &    17.0     &      ---     &    26.8$^{a}$        &  1 \\
H149               &   212.0   &   1.8   &    318.0     &    21.1     &    16.1     &      ---     &    27.2$^{a}$        &  1 \\
H409               &   218.0   &   2.3   &    370.0     &    17.2     &    14.0     &      ---     &    25.7$^{a}$        &  1 \\
H67                &   244.0   &   3.5   &    342.0     &    15.6     &    10.7     &      ---      &    26.3$^{a}$        &  1 \\
N5471-A            &   106.0   &   9.5   &    644.0     &     8.7     &     7.1     &      ---     &    17.5$^{a}$        &  1 \\
N5471-B            &   213.0   &   6.6   &    395.0     &    29.1     &    25.6     &      ---   &    14.5$^{a}$        &  1 \\
N5471-C            &   174.0   &   5.4   &    416.0     &    13.1     &    10.1     &      ---   &    18.8$^{a}$        &  1 \\
N5471-D            &   137.0   &   8.0   &    578.0     &    11.7     &     8.9     &      ---    &    21.6$^{a}$        &  1 \\
N160A1             &   165.5   &   2.4   &    423.6     &    12.7     &    10.1     &      ---    &   37.9        & 2 \\
N160A2             &   164.6   &   1.8   &    382.1     &    11.7     &     9.3     &      ---     &   35.3        & 2 \\
N157B              &   223.0   &   4.8   &    324.0     &    29.1     &    24.5     &      ---     &   35.6        & 2 \\
N79A               &   233.0   &   2.4   &    306.0     &    15.1     &    12.5     &      ---      &   38.3        & 2 \\
N4A                &   152.2   &   2.6   &    430.0     &     9.1     &     6.9     &      ---        &   34.7        & 2 \\
N88A               &    95.6   &   12.0  &    672.0     &     5.2     &     4.5     &      ---       &   17.5        & 2 \\
N66                &   116.9   &   5.6   &    469.0     &    10.1     &     7.0     &      ---       &    20.7        & 2 \\
N81                &   137.0   &   7.4   &    528.0     &     6.9     &     5.2     &      ---        &   18.0        & 2 \\
SDSS\,J1455        &   111.54  &  10.22  &    613.55    &    10.00    &     7.88    &      ---     &    11.54       & 3 \\  
SDSS\,J1509        &   153.18  &   4.20  &    499.42    &    19.66    &    14.88    &      ---     &    25.46       & 3 \\
SDSS\,J1528        &   228.82  &   5.00  &    489.32    &    19.23    &    14.22    &      ---     &    16.88       & 3 \\
SDSS\,J1540        &   217.93  &   2.91  &    309.42    &    26.09    &    19.19    &      ---     &    20.95       & 3 \\
SDSS\,J1616        &    84.91  &   8.51  &    615.16    &     7.74    &     5.81    &      ---       &    16.47       & 3 \\
SDSS\,J1729        &   176.22  &   6.60  &    515.41    &    12.86    &    10.09    &      ---     &    20.92       & 3 \\
1                  &   243     &   3.38  &    321       &    ---      &    ---      &      22.1    &    17.0   & 4 \\
2                  &   166     &   5.59  &    504       &    ---      &    ---      &      17.4    &    19.6   & 4 \\
3                  &   373     &   1.16  &    119       &    ---      &    ---      &      52.0    &    12.7   & 4 \\
4                  &   286     &   2.14  &    253       &    ---      &    ---      &      33.2    &    20.7   & 4 \\
5                  &   296     &   1.10  &    226       &    ---      &    ---      &      52.7    &    26.4   & 4 \\
6                  &   275     &   1.16  &    246       &    ---      &    ---      &      38.6    &    20.3   & 4 \\
8                  &   307     &   0.77  &    144       &    ---      &    ---      &      44.2    &    20.6   & 4 \\
9                  &   172     &   0.89  &    235       &    ---      &    ---      &      33.1    &    27.2   & 4 \\
10                 &   180     &   0.91  &    236       &    ---      &    ---      &      16.3    &   26.5   & 4 \\
11                 &   258     &   1.11  &    201       &    ---      &    ---      &      41.8    &   23.8   & 4 \\
14                 &   248     &   0.75  &    181       &    ---      &    ---      &      41.6    &   29.6   & 4 \\
17                 &   213     &   0.59  &    192       &    ---      &    ---      &      27.3    &   27.3   & 4 \\
19                 &   192     &   0.61  &    165       &    ---      &    ---      &      30.6    &   23.1   & 4 \\
20                 &   146     &   0.71  &    227       &    ---      &    ---      &      20.0    &   28.6   & 4 \\
23                 &   176     &   0.57  &    198       &    ---      &    ---      &      28.0    &   24.0   & 4 \\
24                 &   197     &   0.62  &    180       &    ---      &    ---      &      30.3    &   26.5   & 4 \\
26                 &   160     &   1.19  &    259       &    ---      &    ---      &      22.1    &   28.2   & 4 \\
27                 &   357     &   1.59  &    178       &    ---      &    ---      &      81.7    &   17.3   & 4 \\
28                 &   314     &   1.89  &    244       &    ---      &    ---      &      37.2    &   21.2   & 4 \\
VCC1699      &   126.0   &   4.1   &    553.1     &    10.4   &     7.7    &      ---       &    15.7  & 5 \\
SDSS\,J002101.03   &   163.4   &   5.6   &    433.4     &    13.6     &    10.7     &      ---     &    11.9        & 6 \\   
SDSS\,J003218.60   &   157.3   &   6.2   &    460.7     &    17.3     &    12.5     &      ---     &    21.7        & 6 \\
SDSS\,J162410.11    &  147.1   &   7.0   &    564.2     &    13.6     &     9.9     &      ---     &    16.0$^{a}$  & 6 \\
SDSS J165712.75 (A) &  188.32  &   5.24  &    430.82    &    22.07    &    15.98    &      ---     &    14.00       & 7 \\
SDSS J165712.75 (B) &  132.66  &   8.46  &    486.53    &    14.89    &    10.60    &      ---     &     9.84       & 7 \\    
SDSS J165712.75 (C) &  148.09  &   8.38  &    447.27    &    16.99    &    11.54    &      ---     &    17.77       & 7 \\
SBS\,0335-2013052E  &   20.11  &  11.08  &    327.89    &     1.94    &     1.51    &      ---     &    12.53     &  8 \\
UM\,283D            &  204.65  &   4.10  &    336.63    &    29.30    &    20.80    &      ---     &    20.72 & 9 \\
UM\,133H            &  107.18  &   7.72  &    367.90    &     9.40    &     6.94    &      ---     &      7.42 & 9 \\
HE 2-10C            &  211.87  &   1.96  &    150.52    &    14.69    &    17.87    &      ---     &     34.18 & 9 \\
HE 2-10E            &  193.03  &   0.26  &    126.64    &    14.10    &    15.94    &      ---     &      27.71 & 9 \\
\hline
\end{tabular}
\end{table*}

\begin{table*}
\setcounter{table}{4}
\caption{-$continued$}
\vspace{0.3cm}
\label{tab1}
\begin{tabular}{@{}lccccccccc@{}}
\hline		 
\noalign{\smallskip}  
 Object          &    [\ion{O}{ii}]$\lambda$3727  &   [\ion{O}{iii}]$\lambda$4363       &   [\ion{O}{iii}]$\lambda$5007   &  [\ion{S}{ii}]$\lambda$6717   &   [\ion{S}{ii}]$\lambda$6731 & [\ion{S}{ii}]$\lambda$6725$^{b}$   &[\ion{S}{iii}]$\lambda$9069  & Ref. \\
 \noalign{\smallskip}
NGC\,3125          &    93.12  &   5.13  &    578.95    &    11.97    &     9.92    &      ---          &             31.97  &  9 \\
Mrk\,1259          &   170.87  &   1.08  &    161.78    &    15.75    &    18.13    &      ---          &              23.74  &  9 \\
POX\,4             &    81.26  &   9.12  &    652.36    &     7.63    &     5.94    &      ---              &             12.31  &  9 \\
TOL\,1214-277      &    30.82  &  17.29  &    511.82    &     2.24    &     1.39    &      ---         &              2.85  &  9 \\                            
J\,1253-0312       &    84.22  &  10.37  &    665.63    &     4.94    &     6.25    &      ---           &             13.74  &  9 \\
NGC\,5253 No.C1    &    96.34  &   7.45  &    634.27    &     7.79    &     7.79    &      ---       &             22.44  &  9 \\
NGC\,5253 No.C2    &   148.48  &   3.09  &    445.98    &    14.80    &    12.20    &      ---    &              21.57  &  9 \\
NGC\,5253 No.P2    &   106.83  &   7.92  &    649.56    &     7.19    &     8.00    &      ---      &              22.82  &  9 \\
TOL\,89 No.1       &   151.60  &   3.38  &    489.58    &    11.67    &     8.99    &      ---         &              24.83  &  9 \\
TOL\,89 No.2       &   157.62  &   1.94  &    385.19    &    24.69    &    18.48    &      ---        &              50.12  &  9 \\
TOL\,1457-262      &   258.97  &   6.80  &    627.73    &    11.99    &     9.03    &      ---        &              13.90  &  9 \\
TOL\,1924-426 No.1 &    94.48  &   5.36  &    436.15    &     9.56    &     7.31    &      ---       &             13.07  &  9 \\
TOL\,1924-426 No.2 &   113.90  &   6.23  &    531.57    &    10.88    &     8.59    &      ---     &              13.98  &  9 \\ 
NGC\,6822\,V       &    78.55  &   5.18  &    530.38    &     6.03    &     4.62    &      ---          &              25.60  &  9 \\
VS\,3              &   226.0   &   1.0   &    147.0     &    19.0     &    13.5     &      ---     &    25.1         & 10 \\
VS\,49             &   219.0   &   2.9   &    239.0     &    23.8     &    16.9     &      ---     &   25.5         & 10 \\
VS\,48             &   252.0   &   3.8   &    314.0     &    20.9     &    15.9     &      ---     &   26.8         & 10 \\
NGC\,604           &   215.2   &   0.75  &    207.7     &    16.6     &    11.1     &      ---     &     3.98        & 11 \\
NGC\,588           &   148.2   &   2.4   &    464.7     &    11.8     &     7.7     &      ---     &       5.63        & 11 \\ 
Leo\,P             &    46.5   &   3.8   &    145.3     &     3.6     &     2.7     &      ---     &    14.5            & 12 \\
IC\,132-A          &   150.7   &   6.7   &    500.1     &     8.9     &     5.7     &      ---     &   27.5    & 13 \\                       
IC\,132-B          &   164.7   &   6.4   &    485.6     &     9.9     &     6.6     &      ---     &   24.0    & 13 \\                                
IC\,132-C          &   192.4   &   7.7   &    463.5     &    11.7     &     7.6     &      ---     &  28.4      & 13 \\ 
IC\,132-D          &   210.5   &   8.4   &    449.1     &    12.7     &     8.4     &      ---     &  30.8    & 13 \\ 
IC\,132-E          &   204.5   &  10.7   &    443.5     &    12.3     &     8.2     &      ---     &  30.6     & 13 \\ 
IC\,132-F          &   188.9   &  11.2   &    445.1     &    11.5     &     7.3     &      ---     &  28.1      & 13 \\ 
IC\,132-G          &   221.6   &  14.0   &    435.0     &    12.9     &     8.6     &      ---     & 28.2       & 13 \\ 
IC\,132-H          &   241.3   &  18.9   &    433.3     &    14.0     &     9.2     &      ---     &  23.4      & 13 \\ 
IC\,132-I          &   267.2   &  29.0   &    444.6     &    16.1     &     9.5     &      ---     &    18.9    & 13 \\ 
8                  &   150     &   0.34  &     27       &    44       &    36       &      ---     &   19     & 14 \\
17                 &   143     &   0.40  &    114       &    15       &    11       &      ---     &  23      & 14 \\
25                 &   158     &   1.30  &    251       &    10.9     &     7.9     &      ---     &   21     & 14 \\
26                 &   230     &   1.20  &    220       &    17.7     &    12.4     &      ---     &  22        & 14 \\
35                 &   150     &   1.35  &    320       &    12.7     &     9.2     &      ---     &   21     & 14 \\
32                 &   269     &   0.97  &    102       &    39       &    29       &      ---     &   16       & 14 \\
II\,Zw\,40         &    83.9   &  10.9   &    740.9     &     6.7     &     5.4     &      ---            &      13.9        & 15 \\
Mrk\,5             &   212.9   &   4.4   &    381.5     &    23.3     &    16.6     &      ---           &      15.9       & 15 \\
0749+568           &   166.8   &   9.8   &    488.0     &    17.8     &    11.4     &      ---        &      12.7        & 15 \\
0926+606           &   178.5   &   8.3   &    477.2     &    18.2     &    14.6     &      ---        &      12.2        & 15 \\
Mrk\,709           &   183.6   &   8.8   &    369.6     &    31.3     &    28.5     &      ---         &       8.7      & 15 \\
Mrk\,22            &   148.7   &   8.2   &    545.5     &    11.4     &     8.5     &      ---           &      14.8        & 15 \\
Mrk\,1434          &    96.8   &  10.4   &    502.8     &     9.6     &     6.7     &      ---          &       9.2        & 15 \\
Mrk\,36            &   129.3   &   9.6   &    483.4     &    11.7     &     8.4     &      ---           &      11.9       & 15 \\
VII\,Zw\,403       &   133.3   &   7.1   &    345.5     &    10.3     &     7.5     &      ---         &      11.7        & 15 \\  
UM\,461            &    52.7   &  13.6   &    602.2     &     5.2     &     4.2     &      ---          &      12.4         & 15 \\
UM\,462            &   174.2   &   7.8   &    492.9     &    16.8     &    11.2     &      ---         &      10.5        & 15 \\
Mrk\,209           &    71.9   &  12.7   &    554.3     &     6.1     &     4.5     &      ---           &      12.2        & 15 \\
A                  &   245     &   1.7   &    134.1     &    41.9     &    32.5     &      ---     &    20.8         & 16 \\
N110               &   182     &   1.3   &    140       &    29       &    28       &      ---     &    27.0           & 16 \\
B                  &   242     &   3.1   &    357       &    30.6     &    22.4     &      ---     &     14.2         & 16 \\
C                  &   280     &   1.7   &    257       &    33       &    24       &      ---     &      13.0         & 16 \\
N                  &   173     &   1.2   &    152       &    30       &    29       &      ---     &      31.0        & 16 \\
SE                 &    46.6   &   4.4   &    189       &    ---      &    ---      &       7.2    &     3.9         & 17 \\
NW                 &    26.4   &   6.2   &    185       &    ---      &    ---      &       3.8    &    3.0         & 17 \\
\hline
\end{tabular}
\end{table*}

\begin{table*}
\setcounter{table}{4}
\caption{-$continued$}
\vspace{0.3cm}
\label{tab1}
\begin{tabular}{@{}lccccccccc@{}}
\hline  	
\noalign{\smallskip}  
 Object 	 &    [\ion{O}{ii}]$\lambda$3727  &   [\ion{O}{iii}]$\lambda$4363	&   [\ion{O}{iii}]$\lambda$5007   &  [\ion{S}{ii}]$\lambda$6717   &   [\ion{S}{ii}]$\lambda$6731 & [\ion{S}{ii}]$\lambda$6725$^{b}$	&[\ion{S}{iii}]$\lambda$9069 & Ref. \\
 \noalign{\smallskip}
N12A	 	   &   191     &   3.7	 &    368.4	&     9.2     &	    7.4	    &      --- 	          &    19.9        & 18 \\
N13AB              &   216.3   &   2.7 	 &    347.6	&    13.4     &     8.4	    &      ---	       &    23.6       & 18 \\
N4A	 	   &   203.5   &   2.6	 &    394.4	&    13.4     &    11.0	    &      ---	         &   36.7        & 18 \\
N138A              &   179.1   &   2.1 	 &    371.0	&     8.0     &     6.2	    &      ---	         &   28.0        & 18 \\
Haro 15-B          &    90.5   &   7.8  &     698       &     7.3     &     5.9     &      ---            &   11.9        & 19 \\
 \hline
\end{tabular}
\begin{minipage}[c]{2\columnwidth}
References---	Data compiled by   (1)  \citet{kennicutt03}, (2) \citet{vermeij02}, (3) \citet{hagele08}, 
 (4) \citet{bresolin09}, \\ (5) \citet{vilchez03} (6) \citet{hagele06},  (7) \citet{hagele11}, (8) \citet{izotov06b}, (9) \citet{guseva11},  \\ (10) \citet{garnett97}, 
(11)  \citet{vilchez88}, (12) \citet{skillman13}, (13) \citet{lopezhernandez13}, \\ (14) \citet{zurita12}, (15) \citet{enrique03},
(16) \citet{rosa95}, \\ (17) \citet{skillman93}, (18) \citet{russell90}, (19) \citet{hagele12}. $^{a}$Value computed from \\  $I$([\ion{S}{iii}]$\lambda$9069= 0.4 $\times$ $I$([\ion{S}{iii}]$\lambda$9532).
$^{b}$Sum of the emission-line intensities  [\ion{S}{ii}]$\lambda$6717 and $\lambda$6731.  
 \end{minipage}
\end{table*}

\begin{table*}
\caption{Fluxes  of   infrared emission-lines  (in $\rm W/cm^{2}$) compiled from the literature.}
\vspace{0.3cm}
\label{tab2}
\begin{tabular}{@{}lccccc@{}}
\hline		 
\noalign{\smallskip}
Object          & H\,I\,4.05 $\mu$m & [\ion{S}{iv}]\,10.51$\mu$m  & [\ion{S}{iii}]18.71$\mu$m & Flux $\rm W/cm^{2}$ & Reference \\
\noalign{\smallskip}
IR\,02219       &    6.6    &    26      &   59     &  $ 10^{-18}$  & 19\\
IR\,02219       &    6.7    &    28      &   43     &  $ 10^{-18}$  & 19\\
IR\,10589       &    1.71   &     1.5    &   15.7   &  $ 10^{-18}$  & 19\\
IR\,11143       &    0.68   &     8.4    &    7.8   &  $ 10^{-18}$  & 19\\
IR\,12063       &    2.08   &     9.1    &   11.8   &  $ 10^{-18}$  & 19\\
IR\,12073       &    6.6    &    59      &   35     &  $ 10^{-18}$  & 19\\
IR\,12331       &    0.51   &     1.28   &    9.3   &  $ 10^{-18}$  & 19\\
IR\,15384       &    2.86   &     2.6    &   25.9   &  $ 10^{-18}$  & 19\\
IR\,15502       &    1.5    &     0.23   &    4.9   &  $ 10^{-18}$  & 19\\
IR\,16128       &    0.63   &     0.82   &    6.4   &  $ 10^{-18}$  & 19\\
IR\,17221       &    0.64   &     0.13   &    7.3   &  $ 10^{-18}$  & 19\\
IR\,17455       &    2.0    &     1.9    &   23.2   &  $ 10^{-18}$  & 19\\
IR\,18317       &    2.00   &     0.33   &   25.6   &  $ 10^{-18}$  & 19\\
IR\,18434       &    6.80   &     4.2    &   46     &  $ 10^{-18}$  & 19\\
IR\,18502       &    0.89   &     0.49   &    8.6   &  $ 10^{-18}$  & 19\\
IR\,19207       &    0.67   &     1.41   &    7.4   &  $ 10^{-18}$  & 19\\
IR\,19598       &    4.7    &     3.3    &    7.6   &  $ 10^{-18}$  &         19\\
IR\,21190       &    2.25   &     2.09   &    6.51  &  $ 10^{-18}$  &          19\\
IR\,23030       &    1.60   &     1.3    &   16.3   &  $ 10^{-18}$  &           19\\
IR\,23133       &    1.74   &     0.20   &   10.0   &  $ 10^{-18}$  &           19\\
N160A1          &   51.7    &   289      &  317     &  $ 10^{-20}$  &            20 \\
N160A2          &   55.2    &   194      &  268     &  $ 10^{-20}$  &             20 \\
N159-5          &   12.6    &    56.7    &   74.2   &  $ 10^{-20}$  &             20 \\
N4A             &   10.7    &    75.9    &   93.6   &  $ 10^{-20}$  &             20 \\
N83B            &   12.3    &    18.1    &   28.8   &  $ 10^{-20}$  &             20\\
N157B           &    6.3$^{a}$    &    11.7    &   46.5   &  $ 10^{-20}$  &            20 \\            
N88A            &   14.5$^{a}$    &    55.1   &    8.89  &  $10^{-20}$  &            20    \\
N81            &    9.7$^{a}$    &    15.9   &   14.1   &  $ 10^{-20}$  &      20         \\
NGC3603$^{c}$ &  101.07$^{b}$   &   186.94   &  691.99  &  $ 10^{-20}$  &             21\\
30 Dor$^{c}$  & 168.02$^{b}$    &   811.68   &  802.4   &  $ 10^{-20}$  &             21\\
N66$^{c}$     &   8.754$^{b}$   &    25.72   &   23.44  &  $ 10^{-20}$  &                21\\
638       &   1.37$^{b}$   &      4.94   &    3.09  &  $ 10^{-20}$  &                22\\
623       &   2.98$^{b}$   &      7.45   &    5.04  &  $ 10^{-20}$  &            22\\
45        &   3.55$^{b}$   &     10.1    &   17.4   &  $ 10^{-20}$  &            22\\
214       &   1.39$^{b}$   &  54.2    &    5.18  &  $ 10^{-20}$  &           22\\
33        &   0.901$^{b}$  &   1.22   &    4.91  &  $ 10^{-20}$  &         22\\
42        &   0.625$^{b}$  &   0.573  &    2.56  &  $ 10^{-20}$  &          22\\
32        &   0.657$^{b}$  &   0.238  &    1.48  &  $ 10^{-20}$  &           22\\
251       &   1.00$^{b}$   &   1.57   &    3.02  &  $ 10^{-20}$  &          22\\
301       &   0.915$^{b}$  &   0.475  &    4.69  &  $ 10^{-20}$  &          22\\
4         &   1.78$^{b}$   &   1.46   &    7.80  &  $ 10^{-20}$  &         22\\
79        &   3.00$^{b}$   &   7.70   &   23.1   &  $ 10^{-20}$  &          22\\
87E       &   9.30$^{b}$   &   5.29   &   49.1   &  $ 10^{-21}$  &          22\\
302       &   9.65$^{b}$   &   7.26   &   39.8   &  $ 10^{-21}$  &         22\\
95        &   6.10$^{b}$   &   6.95   &   36.8   &  $ 10^{-21}$  &         22\\
710       &   1.51$^{b}$   &   0.659  &    4.05  &  $ 10^{-20}$  &          22\\
691       &   1.18$^{b}$   &   4.43   &    6.43  &  $ 10^{-20}$  &           22\\
Orion     &   1.448$^{b}$  &   4.765  &   14.49  &  $ 10^{-10}$  &     23\\
G333-North&   2.932$^{b}$  &   0.256  &   14.52  &  $ 10^{-11}$  &      24\\
G333-West &   2.93$^{b}$   &   0.164  &   17.51  &  $ 10^{-11}$  &      24 \\
NGC\,1222   &  15.05$^{b}$   &  22.24   &   64.80  &  $ 10^{-21}$  &           18\\
IC\,342     &  44.16$^{b}$   &   4.76   &  320.03  &  $ 10^{-21}$  &            18\\
NGC\,1614   &  12.56$^{b}$   &   6.89   &   83.03  &  $ 10^{-21}$  &            18\\
NGC\,2146   &  48.35$^{b}$   &   6.30   &  190.12  &  $ 10^{-21}$  &            18\\
NGC\,3256   &  43.94$^{b}$   &   5.25   &  171.83  &  $ 10^{-21}$  &         18\\
NGC\,3310   &   4.11$^{b}$   &   4.46   &   20.50  &  $ 10^{-21}$  &              25\\
NGC\,4676   &   3.40$^{b}$   &   0.56   &   11.32  &  $ 10^{-21}$  &       25\\
NGC\,4818   &  12.14$^{b}$   &   2.21   &   71.96  &  $ 10^{-21}$  &       25\\
NGC\,7714   &  15.47$^{b}$   &  14.76   &   81.50  &  $ 10^{-21}$  &      25\\
\hline
\end{tabular}
\end{table*}

\begin{table*}
\setcounter{table}{5}
\caption{-$continued$}
\vspace{0.3cm}
\label{tab2}
\begin{tabular}{@{}lccccc@{}}
\hline		 
\noalign{\smallskip}
Object                 &     HI\,4.05 $\mu$m   &        [SIV]\,10.51$\mu$m    &      [SIII]18.71$\mu$m &   Flux $\rm W/cm^{2}$  &  Reference \\
\noalign{\smallskip}
W3\,IRS\,5          &       3.5                     &    3.3                        &  16.6                               &        $ 10^{-19}$           &     26\\       
W3\,IRS\,2          &         63.3                 &     228                       & 570                                &        $ 10^{-19}$           &     26\\  
W3\,IRS\,2          &        63.5                  &    259                        & 414                                &        $ 10^{-19}$           &     26\\  
ORION$^{c}$      &       373.9                  &    1116.7                   & 3659.0                           &           $ 10^{-19}$           &     26\\  
Caswell\,H2O\,287.37-\,00.62       &    1.6                        &         2.8                   & 29.8        &           $ 10^{-19}$           &     26\\  
TRUMPLER14     &    6.4                        &   9.6                       &  68.7                                &           $ 10^{-19}$           &     26\\                          
Gal\,287.39-00.63&   5.6                         &   34.0                     &  63.7                                 &           $ 10^{-19}$           &     26\\  
Gal\,289.88-00.79 & 16.3                        &   14.6                     &   149.0                               &           $ 10^{-19}$           &     26\\  
NGC\,3603          &   1.7                         &    15.7                    &   25.4                                 &           $ 10^{-19}$           &     26\\  
RAFGL\,4127       &  6.6                         &    75.7                    &   72.9                                  &           $ 10^{-19}$           &     26\\  
Gal\,298.23-00.33 &  65.9                      &    557.0                   & 382.0                                   &           $ 10^{-19}$           &     26\\  
GRS\,301.11\,+00.97 &  4.7                   &    11.5                     & 77.3                                     &           $ 10^{-19}$           &     26\\  
GRS\,326.44+00.91   &  26.4                 &  25.5                        & 254.0                                   &           $ 10^{-19}$           &     26\\  
15408-5356               & 29.7                 &   53.4                        & 441.0                                     &           $ 10^{-19}$           &     26\\  
G327.3-0.5               &   28.7                &  27.5                         & 497.0                                    &           $ 10^{-19}$           &     26\\  
GRS\,328.30+00.43  &  14.7                 &  2.4                           & 48.8                                       &           $ 10^{-19}$           &     26\\  
15567-5236             &  25.7                   &  6.4                          &  40.4                                       &           $ 10^{-19}$           &     26\\  
16172-5028             &  12.7                 &  4.8                            & 136.0                                         &           $ 10^{-19}$           &     26\\  
G333.13-0.43          &  4.4                   &  12.0                          & 100.0                                           &           $ 10^{-19}$           &     26\\  
Gal\,337.9-00.5       &  15.2                  &  30.0                          &  255.0                                          &           $ 10^{-19}$           &     26\\  
17059-4132            &   17.7                 &      6.7                         & 398.0                                            &           $ 10^{-19}$           &     26\\  
NGC\,6334-A         &   5.0                   &  6.8                             & 44.3                                               &           $ 10^{-19}$           &     26\\
NGC637I               &   11.5                 &   51.0                          &  251.0                                                    &           $ 10^{-19}$           &     26\\                          
Gal\,351.47-00.46   &  5.5                   &   1.5                           &  63.1                                               &           $ 10^{-19}$           &     26\\
NGC\,6357IIIB       &   4.0                  &   9.6                            &   86.4                                                    &           $ 10^{-19}$           &     26\\
RAFGL\,2003        &   46.6                &     5.8                           &   149.0                                               &           $ 10^{-19}$           &     26\\
ARCHFIL$^{c}$     & 13.91                  &  0.9                              &  133.2                                               &           $ 10^{-19}$           &     26\\
Pistol\,star             &  7.1                  &    1.3                             &  32.2                                               &           $ 10^{-19}$           &     26\\
SGR\,D\,H\,II         &  18.8                &   16.8                             &  201.0                                               &           $ 10^{-19}$           &     26\\
RAFGL\,2094         &  11.1                &   1.7                             &  138.0                                               &           $ 10^{-19}$           &     26\\
M\,17 $^{c}$          &  184.23              &   1087.1                        & 2457.2                                                &           $ 10^{-19}$           &     26\\
 18317-0757         &   18.8                &   2.9                             &  260.0                                               &           $ 10^{-19}$           &     26\\
 RAFGL\,2245      &   13.9                &  39.0                             &  357.0                                               &           $ 10^{-19}$           &     26\\
 Gal\,033.91+00.11 &  8.8                &  59.9                             &  5.1                                               &           $ 10^{-19}$           &     26\\
 Gal\,045.45+00.06 & 12.0               &  19.8                             &  119.0                                               &           $ 10^{-19}$           &     26\\
 Gal\,049.20-00.35   & 6.4                & 12.1                              & 76.6                                               &           $ 10^{-19}$           &     26\\
 W51\,IRS2             &  77.4              &  137.0                           &  337.0                                               &           $ 10^{-19}$           &     26\\
IR\,070.29+0160$^{c}$ & 89.8           &  57.6                           &  153.0                                              &           $ 10^{-19}$           &     26\\
S128\,A                     &  2.5              &   4.0                            &  21.0                                              &           $ 10^{-19}$           &     26\\
S138                          &  6.0              & 1.1                              & 44.0                                              &           $ 10^{-19}$           &     26\\
S156\,A                      & 15.5              &   11.8                         &  170.0                                              &           $ 10^{-19}$           &     26\\
S159                          &  16.4            &   1.9                            &  101.0                                              &           $ 10^{-19}$           &     26\\
Hubble\,V                  &  1.9               &   7.5                            &  6.7                                                 &           $ 10^{-20}$           &     27\\
I\,Zw\,36                    &  1.9               &  12                              &  4.9                                                &             $ 10^{-21}$           &     27\\
\hline
\end{tabular}
\begin{minipage}[c]{2\columnwidth}
References---  (19) Data compiled from  \citet{peeters02}, (20) \citet{vermeij02} , (21) \citet{lebouteiller08}, \\
(22) \citet{rubin08},  (23)  \citet{simpson98},  (24) \citet{simpson12},  (25)\citet{bernardsalas09}, \\(26) \citet{giveon02}, and
(27) \citet{nollenberg02}. 
$^{a}$Flux computed from  H\,{\sc i}\,4.051$\mu$m/H\,{\sc i}\,2.63$\mu$m=1.74.\\
$^{b}$Flux computed from  H\,{\sc i}\,4.051$\mu$m/H\,{\sc i}\,12.37$\mu$m=8.2.
$^{c}$Flux obtained assuming the measurements at the different observed positions.  
 \end{minipage}
\end{table*}

\begin{table*}
\caption{Electron temperature, electron density, ionic and total oxygen abundances for the sample
of objetcs   listed in Table~\ref{tab1}. The estimations were  calculated following the Visible-lines method described in  Section~\ref{visib}.
 For the cases where was not possible to calcule the electron density,  a value of $N_{\rm e}=200\: \rm cm^{-3}$ was assumed.}
\vspace{0.3cm}
\label{tab3}
\begin{tabular}{@{}lcccccc@{}}
\hline		 
\noalign{\smallskip}  
 Object            & $T_{\rm e}$ ($10^{4}$) K & $N_{\rm e}$ ($\rm cm^{-3}$) & $\log (\rm O^{+}/H^{+})$ &12+log(O/H) & $\log (\rm S^{+}/H^{+})$ &   $\log(\rm S^{+2}/H^{+})$  \\
\noalign{\smallskip}
H1105    &           0.9046  &  211  &   7.92    &   8.39 &   5.85  & 6.87 \\
H1159    &     	     0.9721 &  ---    &   7.84    &   8.29 &   5.89  & 6.68 \\
H1170    &  	     1.0512  & ---    &   7.93    &   8.16 &   6.10  & 6.85  \\
H1176    &  	     0.9931  & 37    &   7.72    &   8.27 &   5.76  & 6.73 \\
H1216    &  	    1.1277  &   37   &   7.53    &   8.17 &   5.58  & 6.47 \\
H128     &  	     0.9008  &   95    &   7.82    &   8.43 &  5.84   & 6.79 \\
H143     &  	     1.0566  & 58     &   7.77    &   8.16 &  5.95   & 6.59 \\
H149     &  	     0.9582  & 101     &   7.89    &   8.32 &  5.99   & 6.69 \\
H409     &  	     0.9813  & 176   &   7.87    &   8.33 &  5.90   & 6.64  \\
H67        &	     1.1390   &  ---    &   7.73    &   8.13 &  5.71  &  6.51 \\
N5471-A  &         1.3090   &  188  &   7.21    &   8.07 &  5.40  &  6.22 \\
N5471-B  &  	  1.3817   &   301 &   7.46    &   7.91 &  5.91  &  6.09 \\
N5471-C  &         1.2428   &  115    &  7.48    &    8.02 &  5.60  &  6.29 \\
N5471-D	 &         1.2750   &  99    &  7.35    &    8.08 &  5.53  &  6.33 \\
N160A1         &            0.9584  & 147 &  7.78 &   8.38 &   5.78  & 6.81  \\
N160A2         &  	    0.9172  & 145 &  7.84 &   8.41 &   5.78  & 6.82 \\
N157B           &            1.3113  & 232 &  7.53 &   7.93 &   5.93  & 6.51 \\
N79A             &            1.0466 &  200 &  7.81 &   8.20 &   5.80  & 6.73 \\
N4A              &   	      0.9743  & 94   &  7.72  &   8.35 &   5.61  & 6.74 \\
N88A             &  	     1.4242  & 228 &  7.08  &   7.98 &   5.14  & 5.95 \\
N66                & 	      1.2037  & ---   &  7.35  &   8.06 &   5.49  &  6.32  \\ 
N81                & 	      1.2815  & 88   &   7.34 &   8.04 &   5.30  &  6.19 \\
SDSS\,J1455  &   	1.3798  & 142  &   7.18 &   7.99 &   5.42  &  6.04 \\
SDSS\,J1509  & 		1.0688  & 93   &  7.61   &   8.26 &   5.88  &   6.49 \\ 
SDSS\,J1528  &		1.1384  & 67   &   7.70  &   8.22 &   5.82  &  6.30 \\
SDSS\,J1540  & 		1.1073  & 62   &   7.71  &    8.12 &  5.97  &  6.44  \\
SDSS\,J1616  & 		1.2747  & 83   &   7.14  &    8.07 &  5.35  &  6.20 \\
SDSS\,J1729  & 		1.2362  & 136 &  7.49   &    8.10 &  5.60  &  6.31 \\    
1      &             1.1503  & ---  & 7.72  & 8.10  & 5.63  & 6.31  \\	      
2      &             1.1716  & ---  & 7.53  & 8.15  & 5.51  & 6.36  \\      
3      &             1.1205  & ---  & 7.93  & 8.06  & 6.02  & 6.21  \\    
4      &             1.0707  & ---  & 7.87  & 8.17  & 5.86  & 6.46  \\     
5      &             0.9241  & ---  & 8.09  & 8.36  & 6.18  &  6.71 \\      
6      &             0.9174  & ---  & 8.07  & 8.37  & 6.05  &  6.61 \\       
8      &             0.9449  & ---  & 8.07  & 8.26  & 6.08  &  6.58 \\	     
9      &             0.8742  & ---  & 7.93  & 8.35  & 6.02  &  6.79  \\    
10     &            0.8776  & ---  & 7.95  & 8.35  & 5.71   & 6.77  \\   
11     &            0.9524  & ---  & 7.99  &  8.26 &  6.05 &  6.64  \\   
14     &            0.8913  & ---  & 8.06  & 8.33  & 6.10  &  6.80 \\     
17     &            0.8371  & ---  & 8.09  & 8.41  & 5.98  &  6.84 \\      
19     &            0.8698  & ---  & 7.99  & 8.29  & 5.99  &  6.72 \\     
20     &            0.8402  & ---  & 7.93  & 8.38  & 5.84  & 6.85  \\     
23     &            0.8261  & ---  & 8.03  & 8.40  &  6.00 &  6.80 \\	     
24     &            0.8570  & ---  & 8.02  & 8.34  & 6.00  &  6.80  \\     
26     &            0.9120  & ---  & 7.84  & 8.30  & 5.81  &  6.76 \\     
27     &            1.0891  & ---  & 7.95  & 8.14  & 6.24  &  6.37  \\    
28     &            1.0428  & ---  & 7.95  & 8.22  & 5.93  & 6.50  \\  
VCC1699                  &      1.0296  & 69  &  7.57 &  8.34 &    5.62  &  6.49 \\
SDSS\,J002101.03    &  	   1.2406  & 139&   7.46 &  8.03 &   5.62  &   6.03 \\
SDSS\,J003218.60    & 	   1.2607  & 41  &   7.42 &  8.03 &   5.70  &   6.28 \\
SDSS\,J162410.11    &      1.2213  & 50  &   7.43 &  8.13 &   5.62  &   6.16 \\
SDSS J165712.75 (A) & 	 1.2121  & 44  &   7.54  & 8.07 &   5.83  &   6.20 \\
SDSS J165712.75 (B)&   	 1.4069  & 24  &   7.23  & 7.90 &   5.56  &   5.94 \\
SDSS J165712.75 (C) &  	 1.4567  & ---  &   7.25 &  7.85 &   5.59  &   6.00 \\
SBS\,0335-2013052E &  	 2.0093  & 126  &   6.11 &  7.30 &   4.52  &   5.24 \\
UM\,283D             	  &     1.2127  & 24  &   7.58 &  8.01 &   5.95  &   6.35 \\
UM\,133H             	  &     1.5396 &  63  &   7.06 & 7.69  &   5.32  &   5.72 \\
HE 2-10C             	  &     1.2443  &  1082 &  7.57 &  7.81 &   5.75  &   6.55 \\
HE 2-10E             	  &     0.7708  &  723 &  8.19 &  8.44 &   6.10  &   6.94 \\
\hline
\end{tabular}
\end{table*}

\begin{table*}
\setcounter{table}{6}
\caption{-$continued$}
\vspace{0.3cm}
\label{tab3}
\begin{tabular}{@{}lcccccc@{}}
\hline		 
\noalign{\smallskip}  
 Object            & $T_{\rm e}$ & $N_{\rm e}$ & $\log (\rm O^{+}/H^{+})$ &12+log(O/H) & $\log (\rm S^{+}/H^{+})$ &   $\log(\rm S^{+2}/H^{+})$  \\
\noalign{\smallskip}
NGC\,3125                      &      1.0863 &  203  &   7.37  &  8.26  & 5.67  &    6.64  \\
Mrk\,1259             		 &      1.0001 &  841  &   7.74  &  8.05  & 5.92  &    6.59  \\
POX\,4                  	  &      1.2802 &  127  &   7.12  &  8.08  & 5.35  &    6.08 \\
TOL\,1214-277                &      2.0090  &  ---  &   6.30   & 7.49  & 4.54  &     5.14  \\	   
J\,1253-0312        	       &      1.3398  &  1257  &  7.09   &  8.04  & 5.24  &    6.09 \\
NGC\,5253 No.C1  	  &	 1.1964  &  526  &  7.27   &  8.17  & 5.45  &    6.40 \\
NGC\,5253 No.C2   	  &	 1.0102  &  193  &   7.66  &  8.30  & 5.81  &    6.53 \\
NGC\,5253 No.P    	   &	  1.2132  &  789  &   7.30  &  8.16  & 5.43  &    6.39 \\
TOL\,89 No.1        	      &	     1.0093  &   112   &   7.68  &  8.34  & 5.70  &    6.60  \\ 
TOL\,89 No.2        	      &	     0.9315  &   80   &   7.80  &  8.38  & 6.08  &    6.98  \\
TOL\,1457-262      	     &	    1.1618  &  87    &  7.73  &   8.28  & 5.60  &    6.21 \\
TOL\,1924-426 No.1        &      1.2169  &  105 &  7.24  &   8.00  & 5.47  &    6.15 \\
TOL\,1924-426 No.2        &      1.1954  &  143  &  7.34  &   8.11  & 5.55  &    6.19 \\
NGC\,6822\,V                 &      1.1212 &   107 &  7.26 &    8.17  & 5.33 &  6.51 \\
VS\,3                              &      1.0052 &    28 &  7.85 &   8.09  &  5.90 &  6.55 \\   
VS\,49                            &      1.2110 &    31 &  7.61 &   7.94  & 5.86  & 6.36   \\ 
VS\,48                             &      1.2098 &    99 &  7.67 &   8.03  &  5.82  & 6.41\\   
NGC\,604                        &      0.8655 &   ---  & 8.05 &   8.38  & 5.95  & 5.76 \\
NGC\,588                        &      0.9371 &    ---  & 7.77 &   8.43  &  5.73 &  6.04 \\
Leo\,P                              &      1.7296 &    80  &  6.59 &   7.18 &  4.85 &  5.39 \\  
IC\,132-A                         &      1.2584 &     ---  & 7.41 &   8.05  & 5.39  & 6.44  \\		   
IC\,132-B                          &      1.2503 &     ---  & 7.45 &   8.06  & 5.45  & 6.39 	\\		     
IC\,132-C                           &       1.3781 &    ---  & 7.42  &  7.95  & 5.45 &  6.39 \\
IC\,132-D                          &       1.4555 &    ---  & 7.40  &  7.89  & 5.46 &  6.38 \\
IC\,132-E                          &       1.6552 &    ---  & 7.27 &  7.75  & 5.38  & 6.29 \\
IC\,132-F                           &       1.6932 &    ---  & 7.22 &  7.72  & 5.33 &  6.24 \\
IC\,132-G                          &       1.9505 &    ---  & 7.18  & 7.61  & 5.32  & 6.15 \\
IC\,132-H                          &       2.3685 &    ---  & 7.08  &  7.48 &  5.28 &  5.96 \\
IC\,132-I                           &        3.1568 &    ---  & 6.97  & 7.32  & 5.23  & 5.73 \\
8                 &        1.2283 & 190 &  7.43  & 7.51 &  6.14 &  6.30 	 \\
17                &    	 0.8603 & 60   &  7.88  & 8.17 &  5.93  & 6.73 \\
25                &    	 0.9377 & 48   &  7.79  & 8.24 &  5.72  & 6.60 \\
26                &    	 0.9495 & 16  &   7.94  & 8.26 &  5.91 &  6.61 \\
35                &       0.8948 & 48  &   7.84  & 8.38  & 5.82 &  6.65 \\
32                &   	 1.1113 & 73  &   7.80  & 7.95  &  6.14 &  6.32  \\
II\,Zw\,40     &        1.3075 & 171 &  7.11  & 8.11  & 5.28  & 6.05 \\
Mrk\,5         &      	1.1884 & 28  &   7.62 &  8.08 &  5.86 &  6.19 \\
0749+568     &        1.5064 & ---  &   7.27  & 7.85  &  5.58 &  5.97 \\
0926+606    &          1.4070 & 166 &  7.36  & 7.93 &  5.67  & 6.06 \\
Mrk\,709    &            1.6435 & 366 &  7.23 &  7.69 &  5.85  & 5.75 \\
Mrk\,22      &       	 1.3194 & 75   &  7.35 &  8.02 &  5.49  & 6.14 \\
Mrk\,1434     &   1.5286 &  ---  & 7.02 &  7.80 & 5.32  & 5.82  \\ 
Mrk\,36         &   1.4982 &  32  & 7.16 &  7.83 &  5.42 &  5.95 \\
VII\,Zw\,403   &  1.5236  & 47  & 7.16  & 7.70  & 5.36  & 5.93 \\
UM\,461       &    1.5985  & 176 &  6.71 &  7.79 &  5.06 &  5.82 \\ 
UM\,462       &    1.3489  & ---  &  7.40 & 7.98  &   5.63 &  6.06 \\
Mrk\,209       &    1.6105 &  61 &  6.84  & 7.76  & 5.11 &    5.91  \\
A                  &   1.2314 &  122  & 7.64 &  7.84  & 6.11 &     6.38 \\ 
N110            &   1.1027  & 447 &  7.64 &  7.90 &  6.07 &  6.56 \\
B                  &   1.0794 &  57  & 7.79  & 8.21 &  6.06 &  6.29 \\
C                 &   0.9976  & 51  & 7.96  & 8.26 &  6.15  & 6.29 \\
N                 &   1.0486  & 443  & 7.68 &  7.98  & 6.12 &  6.65 \\
SE               &   1.6241  & ---  &  6.65  & 7.33 &  4.94 &  5.25 \\
NW              &   1.9992 & ---.  &  6.24  & 7.09  & 4.56 &  5.06 \\
\hline
\end{tabular}
\end{table*}

\begin{table*}
\setcounter{table}{6}
\caption{-$continued$}
\vspace{0.3cm}
\label{tab3}
\begin{tabular}{@{}lcccccc@{}}
\hline  	
\noalign{\smallskip}  
Object            & $T_{\rm e}$ & $N_{\rm e}$ & $\log (\rm O^{+}/H^{+})$ &12+log(O/H) & $\log (\rm S^{+}/H^{+})$ &   $\log(\rm S^{+2}/H^{+})$  \\ 
 \noalign{\smallskip}
N12A	 &	 1.1318  & 165  &   7.63 &    8.12 &   5.52  &   6.39  \\    
N13AB   &        1.0436 &  ---   &   7.79 &   8.23  &  5.69  &  6.54     \\ 
N4A	 &	  0.9967  & 187  &   7.82 &   8.32  &  5.78  &  6.78   \\ 
N138A    &       0.9582  &  118   &   7.82 &   8.34  &  5.58  &  6.70  \\  
Haro 15-B        1.1748  & 173  &  7.26   & 8.22  & 5.39    &  5.59  \\
 \hline
\end{tabular}
\end{table*}

\begin{table*}
\caption{Ionic  abundances for the sample
of objetcs obtained listed in Table~\ref{tab2}  calculated following the  IR-lines method described in  Section~\ref{corr}
and assuming an electron temperature of 10\,000 K.}
\vspace{0.3cm}
\label{tab4}
\begin{tabular}{lccccc}
\hline		 
\noalign{\smallskip}
Object                     & $12+\log(\rm S^{2+}/H^{+}$)  & $12+\log(\rm S^{3+}/H^{+}$) &    Object                                     & $12+\log(\rm S^{2+}/H^{+}$)  & $12+\log(\rm S^{3+}/H^{+}$) \\
\noalign{\smallskip}  
IR\,02219        &           6.83                            &     5.80                                        &        W3\,IRS\,5                            &         6.56                                     &               5.18                              \\       
IR\,02219        &          6.69                             &    5.82                                         &        W3\,IRS\,2                            &         6.84                                     &               5.76                              \\     
IR\,10589        &           6.85                            &      5.14                                      &        W3\,IRS\,2                             &         6.70                                     &               5.81                              \\     
IR\,11143         &          6.94                             &     6.29                                       &       ORION$^{c}$                          &        6.87                                      &              5.68                               \\    
IR\,12063         &          6.64                              &   5.84                                         &      Caswell\,H2O\,287.37-\,00.62  &        7.15                                    &                5.44                             \\    
IR\,12073          &         6.61                               &  6.15                                          &   TRUMPLER14                          &        6.91                                     &               5.38                              \\   
IR\,12331          &         7.14                               &    5.60                                        &   Gal\,287.39-00.63                      &       6.94                                       &              5.98                               \\   
IR\,15384          &         6.84                               &    5.16                                        &   Gal\,289.88-00.79                      &       6.84                                       &              5.15                               \\   
IR\,15502          &         6.40                               &    4.39                                        &   NGC\,3603                                &       7.06                                       &             6.17                                \\   
IR\,16128          &          6.89                              &     5.32                                       &   RAFGL\,4127                            &       6.93                                       &              6.26                               \\   
IR\,17221          &         6.94                               &     4.51                                       &   Gal\,298.23-00.33                      &       6.65                                       &              6.13                               \\   
IR\,17455          &         6.95                               &     5.18                                       &   GRS\,301.11\,+00.97                 &       7.10                                       &              5.59                               \\   
IR\,18317          &         6.99                               &     4.42                                       &   GRS\,326.44+00.91                   &      6.87                                        &              5.19                               \\   
IR\,18434          &         6.71                               &     4.99                                       &  15408-5356                                &      7.05                                        &             5.46                                \\   
IR\,18502          &         6.87                               &     4.94                                       &   G327.3-0.5                               &        7.12                                      &               5.18                              \\   
IR\,19207          &         6.93                               &     5.52                                       &   GRS\,328.30+00.43                  &        6.40                                      &              4.41                               \\   
IR\,19598          &         6.09                               &     5.05                                       &   15567-5236                               &       6.08                                       &             4.60                                \\   
IR\,21190          &        6.34                                &     5.17                                       &   16172-5028                               &       6.91                                       &             4.78                                \\   
IR\,23030          &        6.89                                &    5.11                                        &  G333.13-0.43                            &        7.24                                      &              5.64                               \\   
IR\,23133          &        6.64                                &    4.26                                        &   Gal\,337.9-00.5                        &        7.11                                      &              5.50                               \\   
N160A1             &       6.67                                &  5.95                                          &  17059-4132                                &       7.24                                       &             4.78                                \\   
N160A2             &       6.57                                &    5.75                                        &  NGC\,6334-A                            &         6.83                                     &              5.33                               \\   
N159-5              &        6.65                                &    5.85                                        &   NGC637I                                &         7.22                                     &              5.85                               \\   
N4A                  &        6.83                                &    6.05                                         &    Gal\,351.47-00.46                  &         6.94                                     &              4.64                               \\   
N83B                &        6.25                                &    5.37                                        &    NGC\,6357IIIB                       &          7.22                                    &                5.58                             \\   
NGC3603          &	6.72				      & 	5.47				      &  RAFGL\,2003 			       &	  6.39				          & 		4.30			    \\        
30 Dor     	     &	   6.56					 &	   5.89				          &   ARCHFIL$^{c}$  			  &  	     6.86			            &		4.01				  \\   
638        	      &  	6.24				   &        5.76			            &    Pistol\,star			          &	    6.54			            & 		4.46			    \\   
623        	      &		6.11			           & 	   5.60			                   &    SGR\,D\,H\,II 			      &  	6.91				          &		5.15				  \\   
45         	       &	6.57				   &	   5.66				            &    RAFGL\,2094			     &	       6.98				        &		4.39				 \\   
214        	      &		6.45			          &	  6.79					   &   M\,17 $^{c}$			      &		7.01				         &		5.97			      \\   
33         	       &	6.62				   &  	  5.33				            &   18317-0757      i			&	     7.02			               & 	     4.39			    \\   
42         	       &	6.50				   & 	  5.16				            &	RAFGL\,2245			      &		7.29				         &		5.65			       \\  
32         	       &	6.24				   &	  4.76				             &	Gal\,033.91+00.11		     &		5.65			               &		6.03			     \\  
251        	      &		6.36			          & 	 5.40				            &  Gal\,045.45+00.06 		     &		6.88			              & 		5.42			    \\  
301        	      &		6.59			          & 	4.92				            &  Gal\,049.20-00.35 		     &		6.96			               & 		5.48			    \\  
4          		&	6.52				    &	 5.12				              &	 W51\,IRS2				&	6.52				           &		 5.45			     \\  
79         	       &	6.77				   &	5.61				             &	     S128\,A                              &	6.81					   &		5.41				 \\  
87E        	      &		6.61			          &  	4.96				            &      S138				         &  	6.75				           &		4.46				  \\  
302        	       &	6.50				  & 	5.08				            &       S156\,A			          &	 6.92					  &		5.08				\\  
95         	       &	6.66			           &	5.26				             &       S159			            &	   6.67				             &		4.27			     \\  
710        	      &		6.31				&	4.84				            &                   				&						   &						 \\  
691        	      &		6.62			        &	5.78				          &	   		                                &						&					      \\  
Orion      	     &		6.88			       &       5.72				     &            			                        &						&					      \\  
G333-North 	 & 	   6.58				   &	   4.14				         & 		 		 &						&					      \\  
G333-West  	 &	  6.66			           &	   3.95				         & 				 &						&					      \\  
NGC\,1222    	 &	  6.52			           &	   5.37				          & 				 &						&					      \\  
IC\,342      	    &         6.74			        &      4.23				     &  				&					       &					     \\  
NGC\,1614    	& 	  6.70				     &     4.94					  &				     &  					    &						  \\  
NGC\,2146    	&	  6.48				     &     4.32					  &				     &  					    &						  \\   
NGC\,3256    	&	  6.48				    &	   4.28				          &  				&					       &					     \\  
NGC\,3310    	&	  6.58				    &	   5.24					 &				    &						   &						 \\  
NGC\,4676    	&	  6.41				    &	   4.42					 &				    &						   &						 \\  
NGC\,4818    	&	  6.66				     &     4.46					  &				     &  					    &						  \\  
NGC\,7714    	&	 6.60				     &	   5.18				           &  				&					       &					     \\  
\hline
\end{tabular}
\end{table*}

\label{lastpage}

\end{document}